\documentclass[usenatbib,a4]{mn2e}
\usepackage{times}
\usepackage{graphicx}

\pubyear{2006}

\author[Bagla and Ray]
{J. S. Bagla$^1$ and Suryadeep Ray$^2$ \\
  $^1$ Harish-Chandra Research Institute,  Chhatnag Road, Jhusi, Allahabad
  211019, India. \\
  $^2$ Inter-University Centre for Astronomy and Astrophysics, Post Bag 4,
  Ganeshkhind, Pune 411007, India \\
  E-mail: $^1$ jasjeet@hri.res.in, $^2$ surya@iucaa.ernet.in}

\title[Gravitational Clustering in Redshift Space]
{Gravitational Clustering in Redshift Space: Non-Gaussian Tail of the
  Cosmological Density Distribution Function}

\label{firstpage}

\def\LaTeX{L\kern-.36em\raise.3ex\hbox{a}\kern-.15em
    T\kern-.1667em\lower.7ex\hbox{E}\kern-.125emX}

\pagerange{\pageref{firstpage}--\pageref{lastpage}} 

\begin{document}

\maketitle


\begin{abstract}
We study the non-Gaussian tail of the probability distribution
function of density in cosmological N-Body simulations for a variety of
initial conditions. 
We compare the behaviour of the non-Gaussian tail in the real space with 
that in the redshift space.  
The form of the PDF in redshift space is of great significance as galaxy
surveys probe this and not the real space analogue predicted using theoretical
models. 
We model the non-Gaussian tail using the halo model.  
In the weakly non-linear regime the moments of counts in cells in the redshift
space approach the values expected from perturbation theory for moments in
real space.  
We show that redshift space distortions in the non-linear regime dominate over
signatures of initial conditions or the cosmological background.
We illustrate this using Skewness and higher moments of counts in cells, as
well as using the form of the non-Gaussian tail of the distribution function. 
We find that at scales smaller than the scale of non-linearity the differences
in Skewness, etc. for different models are very small compared to the
corresponding differences in real space.  
We show that bias also leads to smaller values of higher moments, but the
redshift space distortions are typically the dominant effect. 
\end{abstract}




\section{Introduction}

Large scale structures are thought to have formed by gravitational
amplification of small perturbations 
\citep{1980lssu.book.....P,1999coph.book.....P,2002tagc.book.....P,2002PhR...367....1B}.   
Galaxies are believed to form when gas in highly over dense haloes
cools and collapses to form stars in large numbers
\citep{1953ApJ...118..513H,1977MNRAS.179..541R,1977ApJ...211..638S,1977ApJ...215..483B}.   
Evolution of density perturbations due to gravitational interaction in a
cosmological setting is, therefore, the key process for the study of galaxy
clustering and its evolution. 
It is also one of the most important processes in formation and evolution of
galaxies.   
The basic equations for evolution of density perturbations are well known
\citep{1974A&A....32..391P} and are easy to solve when the amplitude of
perturbations is small.  
Once the amplitude of perturbations at relevant scales becomes large, the
perturbation becomes non-linear and the coupling with perturbations at other
scales cannot be ignored.   
The equation for evolution of density perturbations cannot be solved for
generic perturbations in the non-linear regime. 
One can use quasi-linear approximations for studying mildly non-linear
perturbations
\citep{1970A&A.....5...84Z,1989MNRAS.236..385G,1989RvMP...61..185S,1992MNRAS.259..437M,1993ApJ...418..570B,1994MNRAS.266..227B,1995PhR...262....1S,1996ApJ...471....1H,2002PhR...367....1B}.
Statistical approximations and scaling relations can be used if a limited
amount of information is sufficient
\citep{1977ApJS...34..425D,1991ApJ...374L...1H,1995MNRAS.276L..25J,2000ApJ...531...17Ka,1998ApJ...508L...5M,1994MNRAS.271..976N,1996ApJ...466..604P,1996MNRAS.278L..29P,1996MNRAS.280L..19P,2003MNRAS.341.1311S}. 
In general, however, we require cosmological N-Body simulations
\citep{1998ARA&A..36..599B,2004astro.ph.11043B,1997Prama..49..161B} to follow
the detailed evolution of the system. 

The initial density distribution is often assumed to be a Gaussian random
field \citep{1986ApJ...304...15B}.  
The power spectrum or the two point correlation function is sufficient for
describing such a field.   
Gravitational clustering leads to departures from a strictly Gaussian field
and we need to specify the full hierarchy of $n$-point correlation functions
in order to provide a complete description of the density field at late times 
\citep{1980lssu.book.....P,2002PhR...367....1B}.   
The distribution of matter represented by galaxies, etc. is strongly
non-Gaussian at small scales and hence it is necessary to measure higher order
moments of the galaxy distribution, if not the full probability distribution
function.  

Redshift surveys give us information about the distribution of galaxies in
redshift space.
For a given galaxy, the redshift is a combination of information about the
radial distance and the radial component of the peculiar velocity.  
Although it is possible to construct distribution of galaxies in real space at
large scales, the techniques used are not suitable for recovering information
at small scales \citep{1989ApJ...336L...5B,1990ApJ...364..349D,1992ApJ...391..443N,1997MNRAS.285..793C,1998A&A...335..395B,1998ApJ...508..440N,1999ApJ...515..471N,1999MNRAS.306..491T,1999MNRAS.308..763M,2000ApJ...533...50W,2002Natur.417..260F,2002MNRAS.335...53B,2003MNRAS.346..501B,2003A&A...406..393M,2004astro.ph.10063M}. 
Therefore we are confined to studying the galaxy distribution in redshift
space at small scales, and hence it is important to understand the effect of
redshift space distortions on the distribution of galaxies.

It is well known that galaxies do not trace the matter distribution, i.e.,
the galaxy distribution is biased. 
This is reflected in the clustering properties of galaxies which should differ
from the clustering of matter, e.g., the correlation function and the
reduced moments are expected to be different for galaxies.   
It is known that highly over dense regions cluster much more strongly than the
overall distribution of matter, and the peak picture provides a natural
explanation for this \citep{1984ApJ...284L...9K}.  
This can explain the large two point correlation function of galaxies at high
redshifts, e.g. see \citet{1994ApJ...431..477B,1998MNRAS.297..251B}. 
It can be combined with other inputs to develop an understanding of the
evolution of galaxy clustering 
\citep{1996MNRAS.282..347M,1996ApJ...461L..65F,1997MNRAS.286..115M,1998MNRAS.299..417B,1999MNRAS.305..151R}.

The clustering of galaxies depends on the environment, galaxy type and other
factors that are relevant for galaxy formation and evolution.  
It is difficult to codify the physics of galaxy formation in simple terms and
hence these aspects are modelled in terms of a stochastic bias
\citep{1998ApJ...500L..79T,1999ApJ...520...24D,2001MNRAS.323....1S,2004MNRAS.350.1385S}.  

In this paper we study the counts in cells using N-Body simulations.  
We model the probability distribution function in redshift space using the
halo model. 
We also study the effect of bias on moments of
counts in cells, both in real and redshift space.   

The paper is organised as follows: we review some basics of gravitational
clustering in \S{2}.  
In \S{3} we discuss the simulations used in this work and also present the key
results.  
We conclude with a discussion (\S{4}).

\section{Gravitational Clustering}

The evolution of density perturbations can be studied in the Newtonian limit
at sufficiently small scales for non-relativistic matter. 
Indeed, it has been shown that the Newtonian analysis gives the correct
solution up to at least the third order for cold dark matter models even when
a finite cosmological constant is present \citep{2005astro.ph..7159H}.  
Thus the Newtonian limit can be used with confidence at scales
sufficiently smaller than the Hubble radius. 
Evolution of perturbations can then be described by the following equations. 
\begin{eqnarray}
  \ddot{\mathbf  x}_i +  2 \frac{\dot{a}}{a} \dot{\mathbf  x}_i
  &=& - \frac{1}{a^2} {\mathbf \nabla}_i ~ \varphi      \nonumber \\
\nabla^2\varphi = 4 \pi G a^2 \left(\rho - \bar\rho\right) &=&    \frac{3}{2}
  H_0^2 \Omega_{nr}  \frac{\delta}{a}    \nonumber \\ 
\rho({\mathbf x}) &=& \sum\limits_i m_i ~ \delta_{_D}({\mathbf x} - {\mathbf x}_i) 
\end{eqnarray}
It is assumed that the density field is generated by a distribution of
particles, each of mass $m_i$ and comoving position ${\mathbf x}_i$.
$H_0$ is the present value of Hubble constant, $\Omega_{nr}$ is the present
density parameter for non-relativistic matter and $a$ is the scale factor. 
Density contrast $\delta = ( \rho - \bar\rho ) / \bar\rho$, where $\bar\rho$
is the average density.  
It is clear from these equations that over dense regions $(\delta > 0)$ grow
by pulling in matter from surrounding regions while under dense regions
$(\delta < 0)$ become more under dense and start to occupy larger volume. 

It can be shown that gravitational instability leads to formation of highly
dense haloes that are in virial equilibrium or close to it 
\citep{1974A&A....32..391P,1980lssu.book.....P}.  
These haloes are the dominant structures at late times and several
properties of the density field can be deduced by considering the
haloes to be the basic unit
\citep{1996MNRAS.282..347M,1997MNRAS.284..189M,2002PhR...372....1C}. 
This approach is known as the halo model. 

\subsection{Moments of Counts in Cells}

Given a distribution of points, estimation of $n$-point correlation functions
is a very difficult task as this computation requires $\mathcal{O}(N^n)$
floating point operations for a set of $N$ particles.  
One can, however, compute the moments of counts in cells and recover some
information about $n$-point correlation functions
\citep{1980lssu.book.....P,1995ApJS...96..401C}.    
In this method, we distribute a large number of cells in the volume containing
the distribution of points and count the number of particles in each cell. 
The count probability distribution function (CPDF) is a function of the cell
size and the count, of course.  
The CPDF contains full information about the distribution of points.  
It can be used to estimate the moments, which in turn are related to the
$n$-point correlation functions.   
The number of operations required for calculating the CPDF at a given scale is
$\mathcal{O}(N \bar{n}V)$, where $\bar{n}$ is the number density of particles
and $V$ is the volume of each cell.  
This reduction in the computation required is remarkable but there is also a
reduction in the amount of information we can recover about the $n$-point
correlation functions.   
To illustrate this, we give some basic definitions here: 
\begin{eqnarray}
{\bar{N}}_c &=& \langle N_c \rangle ~ = ~ {\bar n} V \nonumber \\
\mu_j &=& \langle \left( N_c - {\bar{N}}_c \right)^j \rangle ~~~~ ; ~~~~
j=2,3,... \label{moments1}
\end{eqnarray}
where $N_c$ is the number of particles in a cell and ${\bar{N}}_c$ is its
average value for cells of a given size.  
$\mu_j$ are moments of the CPDF about the average and these can be related to
$n$-point correlation functions.
\begin{eqnarray}
\frac{\mu_2}{{\bar{N}}_c^2} &=&   \bar{\xi}(r) + \frac{1}{{\bar{N}}_c} 
\nonumber \\
\frac{\mu_3}{{\bar{N}}_c^3} &=&  \bar\zeta(r) + 3
  \frac{\bar{\xi}(r)}{{\bar{N}}_c} +  \frac{1}{{\bar{N}}_c^2} 
\nonumber \\ 
\frac{\mu_4}{{\bar{N}}_c^4} &=&   \bar\eta(r) 
+ 3 \bar{\xi}^2(r)
+ 6 \frac{\bar\zeta(r)}{{\bar{N}}_c}   \nonumber \\
&& +  \left( \frac{7}{{\bar{N}}_c^2} + \frac{6}{{\bar{N}}_c} \right)
\bar{\xi}(r) 
+  \frac{3}{{\bar{N}}_c^2}
-  \frac{4}{{\bar{N}}_c^3} 
\label{moments2}
\end{eqnarray}
Here, $\bar\xi$ is the mean two point correlation function, $\bar\zeta$ is the
mean three point correlation function and $\bar\eta$ is the corresponding four
point function \citep{1980lssu.book.....P}.  
Note that lower order correlation functions contribute a non-trivial amount to
higher order moments and $\bar\xi$, $\bar\zeta$, $\bar\eta$, etc. correspond
to the irreducible part of the $n$th order correlation function. 
These expressions include corrections for shot noise. 
From here, we can define reduced moments as:
\begin{eqnarray}
S_3 &=& \frac{\bar\zeta(r)}{\bar{\xi}^2(r)} \nonumber \\
S_4 &=& \frac{\bar\eta(r)}{\bar{\xi}^3(r)} ~~~~~ .
\label{sq}
\end{eqnarray}
Here, $S_3$ is the Skewness and $S_4$ is Kurtosis. 
These reduced moments are expected to vanish in the linear regime for Gaussian
initial conditions. 
Gravitational instability introduces mode coupling and this leads to
generation of non-Gaussian features.
Perturbation theory can be used to estimate the values of Skewness and other
reduced moments in the weakly non-linear regime ($\bar\xi \ll 1$).
For power law models in an Einstein-de Sitter universe, the reduced moments do
not depend on scale in this regime
\citep{1980lssu.book.....P,1996ApJ...465...14C}.  
\begin{eqnarray}
S_3 &=& \frac{34}{7} - (n + 3) \nonumber \\
S_4 &=& \frac{60712}{1323} - \frac{62}{3} (n + 3) + \frac{7}{3} (n + 3)^3
\label{sn_lin}
\end{eqnarray}
here $n$ is the index of the initial power spectrum of density perturbations.  
It is difficult to obtain a direct solution in the strongly non-linear
regime. 
Stable clustering and hierarchical clustering predict a constant asymptotic
value \citep{1977ApJS...34..425D,1996ApJ...465...14C}
\begin{equation}
S_3 = \frac{9}{n + 3}
\label{sn_nl}
\end{equation}
Therefore we expect $S_3$ (and also $S_4$) to grow from their value in the
weakly non-linear regime (Eqn.(\ref{sn_lin})) to a larger value in the
extremely non-linear regime (Eqn.(\ref{sn_nl})).  
The increase is larger for indices with small $n+3$, as is clear from
eqn.(\ref{sn_nl}). 
Small departures from the stable clustering hypothesis have been seen in
simulations \citep{1996ApJ...465...14C}.

\subsection{Redshift Space Distortions} 

Redshift surveys of galaxies give us information about the location of
galaxies in the redshift space.
Combination of the peculiar motion of galaxies and cosmological expansion
determines the redshift of a galaxy and as a result the location in
redshift space is different from the location of the galaxy in real space. 
This affects the estimated position along the line of sight and does
not change the location transverse to the line of sight.  
Thus redshift space distortions introduce anisotropy in the distribution of
galaxies, even if there is no statistical anisotropy in the galaxy
distribution in real space. 
On small scales, the random motions in virialised haloes stretch these into
radially elongated {\sl fingers-of-god}. 
Even though the effect originates from random motions in compact clusters, it
is relevant at scales as much as twenty times the size of a typical cluster. 
On very large scales, infall due to gravitational clustering compresses
over dense regions along the line of sight. 
The compression is a function of infall velocities, and these depend
on the density parameter of non-relativistic matter.
This compression leads to an amplification of power spectrum and
correlation function at these scales \citep{1987MNRAS.227....1K}.  
For galaxies and other tracers of the matter distribution, the compression
also depends on bias, e.g., see \citet{1999ApJ...520...24D}.

The halo model 
\citep{1996MNRAS.282..347M,1997MNRAS.284..189M,2002PhR...372....1C} has been 
used to model the effect of redshift space distortions on the non-linear power
spectrum at scales where the dominant contribution is from random motions in
virialised haloes \citep{2001MNRAS.321....1W,2001MNRAS.325.1359S}. 
At small scales, where the coherent motion of galaxies due to large
scale flows \citep{1987MNRAS.227....1K} can be ignored, the relation between 
the observed density in real space $\rho^{(r)}({\mathbf r})$ and that in 
redshift space $\rho^{(s)}({\mathbf s})$ can be expressed as follows 
\citep{1994ApJ...424...30M}.
\begin{equation}
\rho^{(s)}({\mathbf s}) = \int\limits_{-\infty}^{+\infty} dv f_{\sigma}(v) 
\rho^{(r)}({\mathbf s} + H_0^{-1} v \hat{\mathbf s}) ~~ 
\label{matsub1}
\end{equation}
where $f_{\sigma}(v)$ is the distribution function of peculiar velocities and 
\begin{equation}
\int\limits_{-\infty}^{+\infty} dv f_{\sigma}(v) = 1 ~~~ .
\end{equation}
In these equations $H_0$ is, as before, the present value of Hubble constant.
The observer is presumed to be at the origin of the coordinates and 
$\hat{\mathbf s} \equiv {\mathbf s}/|s|$ is the unit vector towards
any location ${\mathbf s}$ in the redshift space. 

Within the formalism of the halo model the amplitude of
the power spectrum at small scales (large $k$) is dominated by halo profiles
and virial motions within haloes act to reduce the amplitude of the power in
redshift space.  
If one assumes that in real space the halos are isotropic, virialised
and approximately isothermal with $1D$ velocity dispersion $\sigma$, then the
peculiar motions {\it within\/} a halo add a Gaussian noise to the
redshift space radial coordinate \citep{2001MNRAS.321....1W} in Fourier
space. 
\begin{equation}
\rho^{(s)}_k = \rho^{(r)}_k \exp\left[-\frac{( k H_0^{-1} \sigma
    \mu)^2}{2}\right], \label{white1}
\end{equation}
where $\mu=\hat{r}\cdot\hat{k}$.  
The effect of redshift space distortions on the power spectrum can be
calculated by integrating over $\mu$ \citep{2001MNRAS.321....1W}.  
However, we are interested in higher moments and the anisotropic halo
profile is required for an accurate description.


\begin{table*}
\begin{center}
\begin{tabular}{||l|l|l|l|l|l||}
\hline
\hline
Model & $N_p$ & $N_{cell}$ & $L_{box}$ & $\epsilon$ & $r_{nl}$ \\
\hline
\hline
Power Law $n=-2.0$ & $256^3$ & $256^3$ & $256$ (grid units) & $0.4$ (grid
units) & $5.0$ (grid units) \\
\hline
Power Law $n=-1.5$ & $256^3$ & $256^3$ & $256$ (grid units) & $0.4$ (grid
units) & $8.0$ (grid units) \\ 
\hline
Power Law $n=-1.0$ & $256^3$ & $256^3$ & $256$ (grid units) & $0.4$ (grid
units) & $8.0$ (grid units) \\ 
\hline
Power Law $n=-0.5$ & $256^3$ & $256^3$ & $256$ (grid units) & $0.4$ (grid
units) & $9.0$ (grid units) \\ 
\hline
Power Law  $n=0.0$ & $256^3$ & $256^3$ & $256$ (grid units) & $0.4$ (grid
units) & $10.0$ (grid units) \\ 
\hline
$\Lambda$CDM & $256^3$ & $256^3$ & $200$h$^{-1}$Mpc & $0.47$h$^{-1}$Mpc  
& $8.0$h$^{-1}$Mpc \\
\hline
\hline
\end{tabular}
\caption{This table lists the fixable parameters of the N-Body simulations we
  have used.  All the simulations were done using the TreePM code; the
  detailed configuration is described in \citet{2003NewA....8..665B}.  The
  first column lists the model used in generating the initial conditions for a
  simulation, the second column lists the number of particles in the
  simulation, the third column the size of the cubic grid used within the
  simulation volume to compute Fourier transforms for the long range part of
  the force of gravitational interaction, the fourth column the length of a
  side of the simulation box in relevant units and the last column the
  softening length for force in the same units. A {\it grid unit} here is
  defined as the length of the side of the smallest cell in the grid within
  the simulation box. The last column lists the largest scale of non-linearity 
  $r_{nl}$ for a given model that we have used for our analysis.} 
\end{center}
\end{table*}


\section{N-Body Simulations}

In the present work, we study the moments of counts in cells in N-Body
simulations.  
We compute $\bar\xi$, $S_3$ and $S_4$ for a variety of initial conditions in
real space as well as redshift space.  
We primarily concentrate on simulations of power law models in an
Einstein-de Sitter background Universe ($\Omega_m = 1.0$): 
\begin{equation}
P(k) = A k^n
\end{equation}
We ran simulations of power-law models with index $n = - 2$, $- 1$ and
$0$. 
These were normalised such that linear fluctuations at the scale of $8$
grid lengths were unity when the scale factor $a = 1$. 
Power law models do not have any intrinsic scale except for the scale of
non-linearity introduced by gravity. 
Self similar evolution of indicators of clustering is a good check for the
accuracy of simulations for such models.   

We also simulated the $\Lambda$CDM model with $\Omega_b = 0.05$,
$\Omega_{dm}=0.25$, $\Omega_{tot}=1.0$, $h=0.7$ and $n=1.0$, where $h$ is the
dimensionless Hubble parameter and $n$ is the primordial spectral
index. 
The power spectrum was normalised by choosing $\sigma_8(z=0)=1$.

The parallel TreePM method
\citep{2002JApA...23..185B,2003NewA....8..665B,2004astro.ph..5220Ra} was used
for all simulations.  
We assumed that the particles have a finite size given by a spline kernel
\citep{2001NewA....6...79S} to obtained a softened force. 
The softening length used is listed in table~1 along with other parameters of
simulations used here.  

We also study clustering properties of over dense regions as this approximates
the distribution of galaxies, though it is more appropriate to work
with a model for occupation number for a more refined study
\citep{2001MNRAS.325.1039B}. 
We identify high density regions in our simulations by using a density
cutoff. 
We compute the Lagrangian density of particles in a given simulation
using a top-hat kernel with a radius equal to the distance to an $n$th
neighbour with $n \simeq 10$.
We then retain particles in regions with density greater than a given
threshold. 

We use power law models as the requirement of scale invariant evolution is a
powerful constraint and it can be used to ensure that the simulations do not
suffer from any serious defect. 
This is a particularly useful approach for models with a lot of power at large
scales as in this case the effects of a finite box size can be important
\citep{2005MNRAS.358.1076B,2006astro.ph.01320B}. 
We find that in the simulations we have used here, scale invariance is
satisfied up to the last epoch used here for the second moment.  

In order to calculate the CPDF, we used a large number of uniformly
distributed spherical cells. 
The number of cells varied from $10^8$ for the smallest cells to $1.6 \times
10^7$ for the largest cells.
Radii of cells were varied in the range $0.6$ to $20$ grid lengths. 
As the largest cell size used is much smaller than the simulation box
size, cosmic variance is not a concern. 

\begin{figure}
\begin{center}
\includegraphics[width=2.5in]{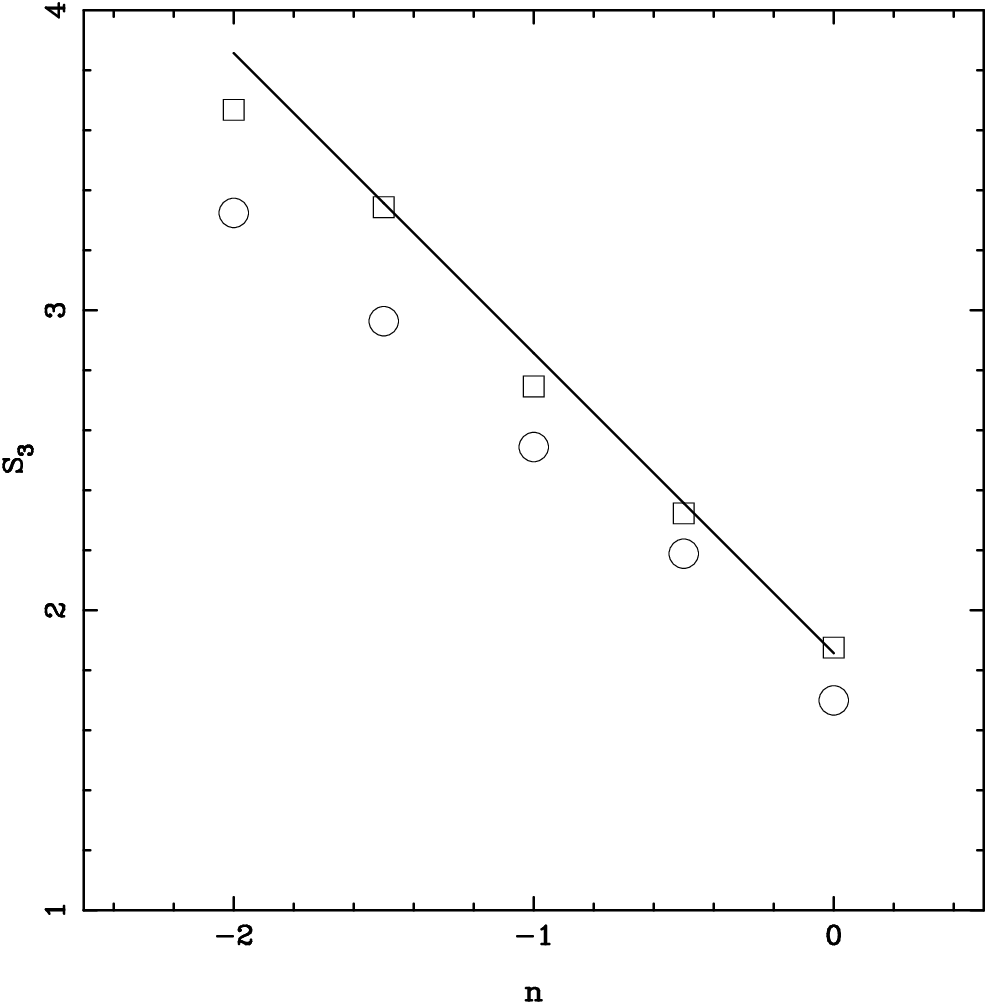}
\includegraphics[width=2.5in]{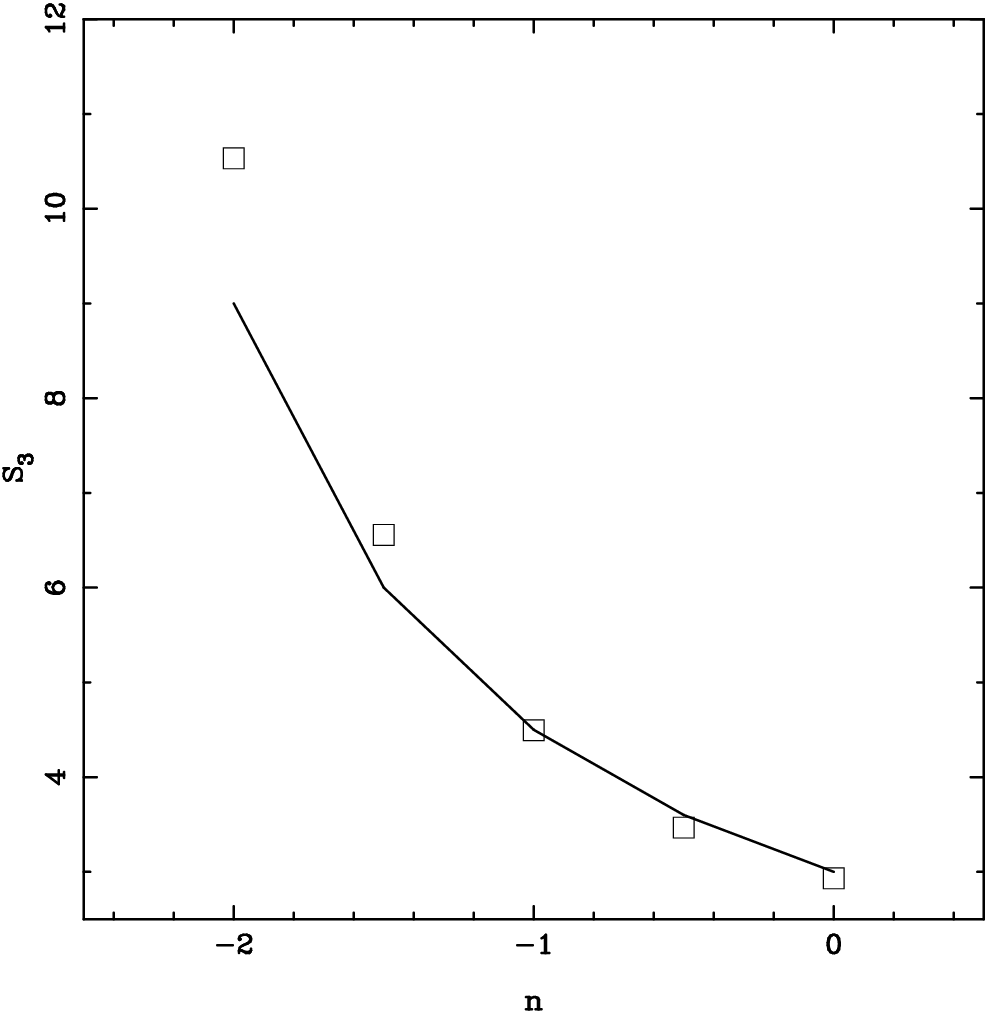}
\end{center}
\caption{$S_3$ has been plotted for power law models as a function of the
   power law index $n$ in the weakly non-linear regime for both real and
   redshift space in the upper panel.  For a given $n$, we have plotted the
   average $S_3$ with $0.1 \leq \bar\xi \leq 0.2$.  We used data from several
   epochs for each model.  The average $S_3$ is plotted as a function of $n$.
   The rectangles correspond to the value of $S_3$ in real space, whereas the
   circles show the same in redshift space. The full line represents the
   predicted values from perturbation theory (refer to the text for further
   details).  In the lower panel $S_3$ has been plotted as a function of $n$
   in real space in the non-linear regime ($\bar\xi \geq 30$).  The curve
   shows the predicted variation in the stable clustering hypothesis.} 
\end{figure}

We first verify the perturbative prediction for $S_3$ in the weakly non-linear
regime, and also check for any variations in moments in this regime as we go
to redshift space. 
Figure~1 shows $S_3$ as a function of the spectral index $n$ for power law
models.  
This value of $S_3$ is the average obtained from scales where $0.1 \leq
\bar\xi \leq 0.2$.  
We do not expect the values of $S_3$ to deviate too strongly from the
perturbative predictions at these scales. 
As seen in the figure, the values of $S_3$ are close to the predictions given
in Eqn.(\ref{sn_lin}). 
The values of $S_3$ encountered in redshift space are generally smaller,
mainly due to the fingers of god effect. 
For analytical predictions of Skewness in redshift space see
\citet{2000MNRAS.314...92T,2001MNRAS.320..139W}. 

We also study the variation of $S_3$ in the strongly non-linear regime. 
The lower panel of Figure~1 shows $S_3$ as a function of $n$ for $\bar\xi \geq
30$.  
Also shown is the variation expected from the stable clustering hypothesis
(Eqn.(\ref{sn_nl})). 
As stable clustering is expected only at much higher levels of
non-linearity, we expect these simulations to under-estimate $S_3$ as
compared to the analytical prediction.  
The simulations and the analytical predictions are consistent with each other
at this level, except $n=-2$ where we believe the effects of a finite box size
are starting to manifest themselves
\citep{2005MNRAS.358.1076B,2006astro.ph.01320B}.  
The real test of the stable clustering hypothesis requires carrying out this
comparison at $\bar\xi \gg 100$ and we have not reached such non-linearities
in the simulations used here. 

\begin{figure}
\begin{center}
\includegraphics[width=2.5in]{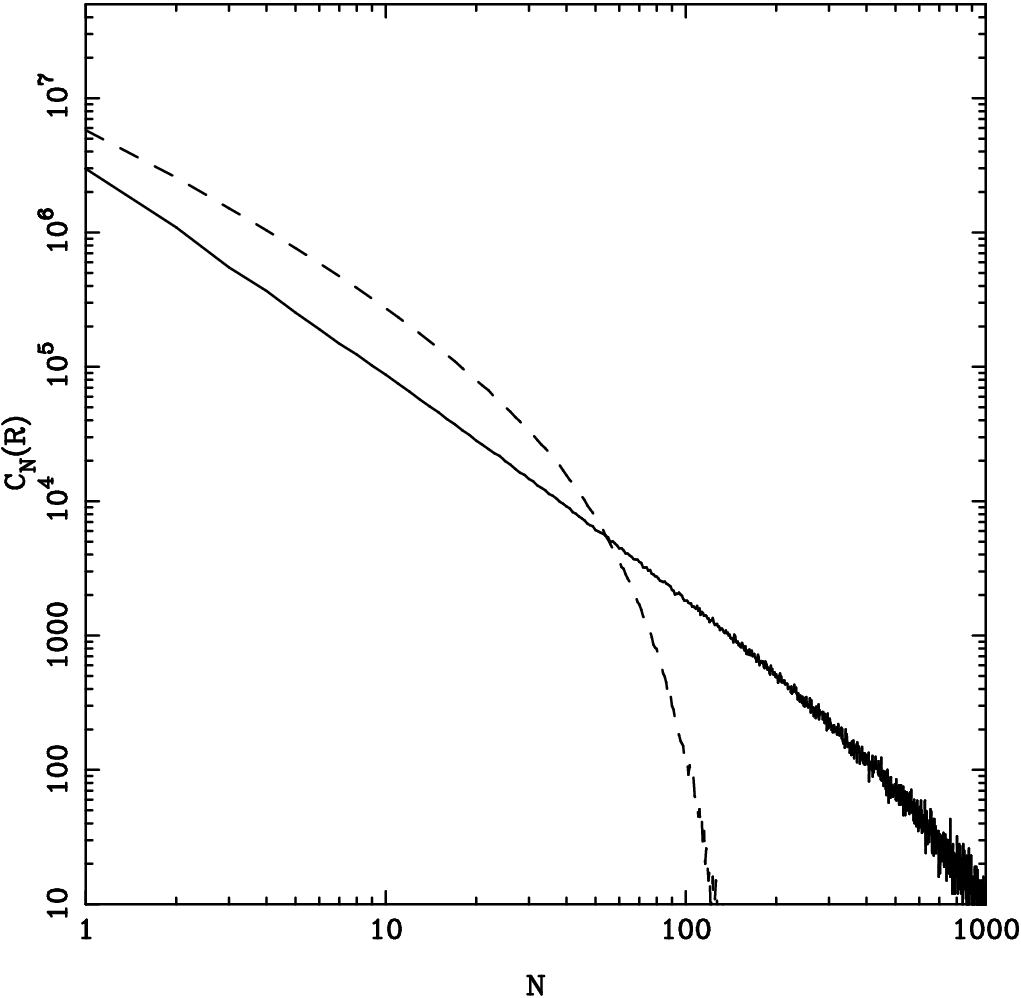}
\includegraphics[width=2.5in]{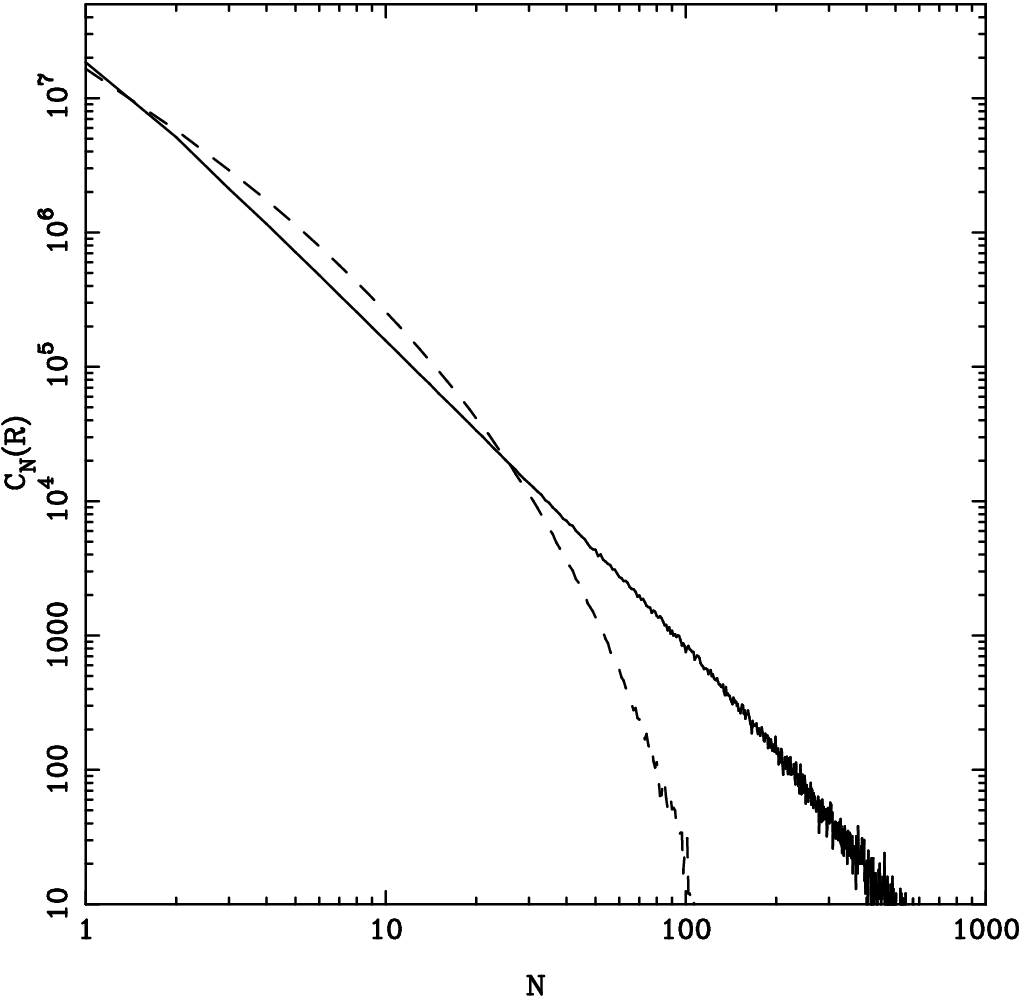}
\end{center}
\caption{The distribution of counts $C_N$ in cells is shown for power law
  models with $n=0$ (top panel) and $n=-2$ (lower panel). These are for cell
   of size of $R=0.6$ grid lengths. This scale is smaller than the
  virial radius of largest haloes in the simulations.  Solid line 
   shows the counts in real space and the dashed line shows the counts in
   redshift space.  The plots were made using data for the highest $r_{nl}$
   listed in table~1 for the corresponding model.}  
\end{figure}

Having verified known results about the moments, we proceed with a discussion
of the non-Gaussian tail of the CPDF.  
We start with a comparison of the tail of CPDF with an analytical
calculations done with the halo model. 
Figure~2 shows the CPDF for the $n=0$ and $n=-2$ power law models for a small
cell-size ($l=0.6$~grid length) in both real and redshift space. 
We find that the tail of the CPDF in redshift space falls off much more
sharply as compared to the corresponding tail of CPDF in real space.  
Partly due to the different slope, the CPDF in redshift space is larger than
the CPDF in real space at intermediate densities. 
The basic features are similar for $n=0$ and $n=-2$ and this suggests that
redshift space distortions may be responsible for the key features rather
than any power spectrum dependent quantity. 
The differences in the two panels here may be attributed to the
different scale of non-linearity, as this results in different sizes
for the most massive haloes whereas the cell size used here is the
same for both the models. 

\begin{figure}
\begin{center}
\includegraphics[width=2.5in]{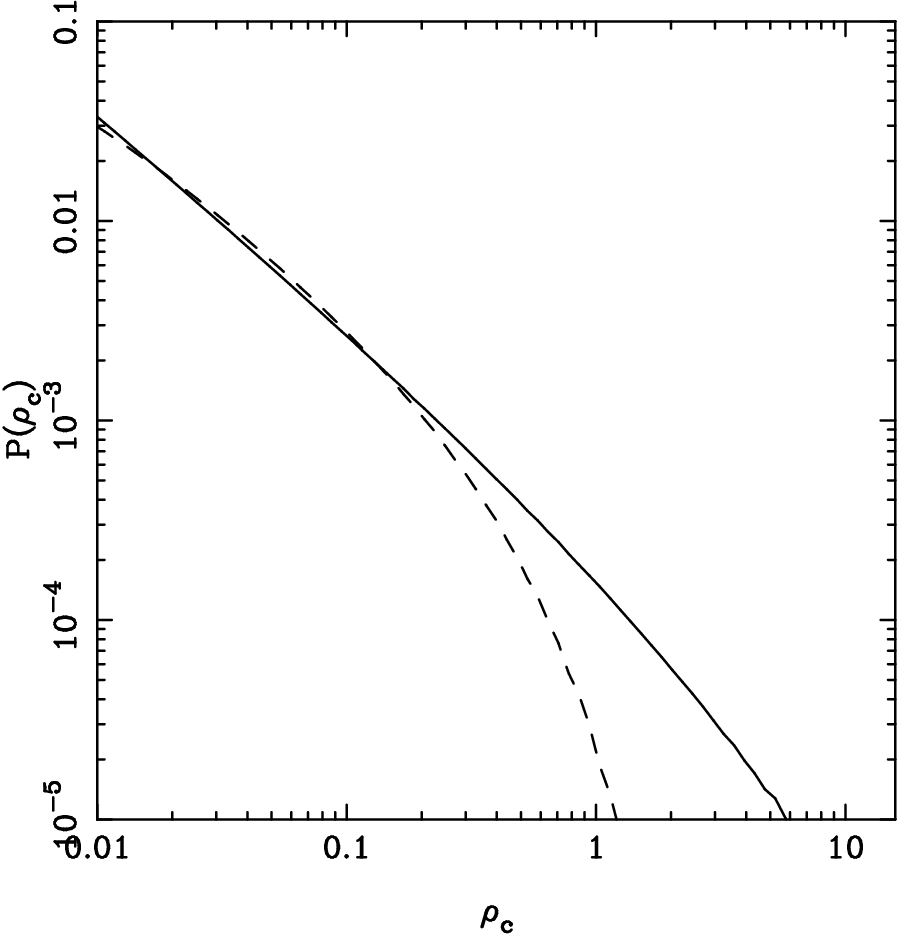}
\includegraphics[width=2.5in]{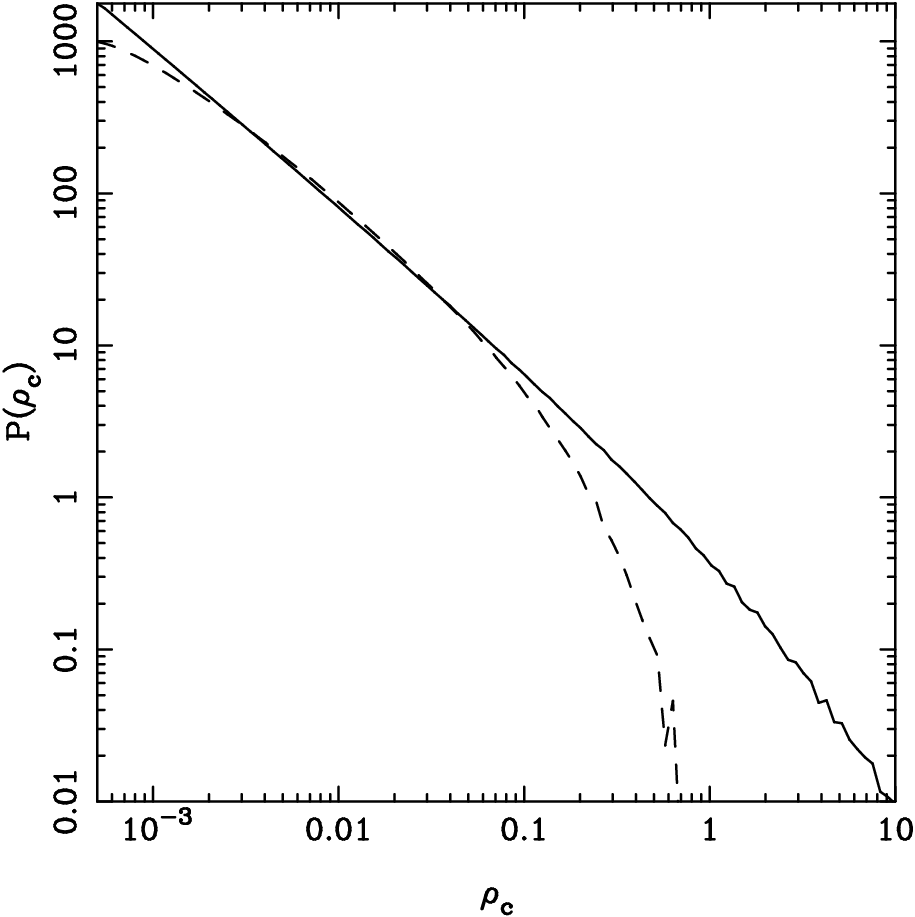}
\end{center}
\caption{The upper panel shows the unnormalised probability distribution
  function for a NFW halo plotted as a function of average density in the
  cell $\rho_c$.  A cell size of $0.01~r_{vir}$, and a uniform $\sigma$
  evaluated at the virial radius was used here. Density is shown in units of
  the density at the virial radius. The solid line shows the distribution
  function in real space and the dashed line is for the distribution function
  in redshift space.  This brings out the key feature of a cutoff in the
  probability distribution in redshift space.  In contrast, the real
  space distribution 
  function tapers off at much higher densities.  The lower panel shows the
  volume occupied by regions with density $\rho_c$ as a function of $\rho_c$
  for the $n=-2$ power law model.  Both the density and volume are in
  arbitrary but self consistent units here.  The volume is equivalent to
  unnormalised probability.  The solid line is for real space
  and the dashed line for redshift space.  These have been computed using the
  halo model with the Press-Schechter mass function and assuming a NFW
  profile.  Refer to the text for further details.}  
\end{figure}

\subsection{The Non-Gaussian Tail}

Higher moments are sensitive to the form of the non-Gaussian tail of the CPDF
and hence we need to model the contribution of central parts of haloes to the
CPDF. 
We model the non-Gaussian tail of the CPDF in redshift space using the
halo model.  
We follow the approach used by \citet{1994A&A...281..301C,2003MNRAS.339..495T}
for this, and extend their formalism to redshift space. 

Let us assume that in real space the haloes are spherical with a central
profile that can be approximated as 
\begin{equation}
\rho^{(r)}(r) = \rho_c \left(\frac{r}{r_c}\right)^{-\alpha}
\label{colombi1}
\end{equation}
As the density field is represented by point particles, the number density
of these particles is proportional to the density $\rho$.  
If the cell size is much smaller than characteristic scales of typical haloes
then the tail of the CPDF can be estimated approximately by computing the
volume occupied by regions with density greater than a threshold.  
As long as the effects of a finite number of particles are not important, the
CPDF is expected to have the form $P_N(R) \propto N^{-3/\alpha - 1}$, with $R$
being the cell size \citep{1994A&A...281..301C}. 
Use of a finite sized cell implies smoothing of the density field
\citep{2003MNRAS.339..495T}, therefore the highest density sampled in a given
halo is finite even if the actual density profile is singular.   
The smoothing also implies that there is a minimum halo mass and only haloes
more massive than this minimum contribute to the non-Gaussian tail, though
this cutoff is rarely of much significance as much of the contribution to the
tail of the CPDF comes from most massive haloes.  
To generalise the calculation of the form of the non-Gaussian tail to haloes
in redshift space, we note that in this case the density profile
cannot be assumed to be isotropic.    
We first compute the density profile in redshift space (Eqn.\ref{matsub1}) and
then compute the volume ($V(>\rho)$)occupied by regions with density above a
threshold.     
We may approximate Eqn.(\ref{matsub1}) as:
\begin{eqnarray}
\rho^{(s)}({\mathbf s}) &=& \int\limits_{-\infty}^{+\infty} dv f_{\sigma}(v) 
\rho^{(r)}({\mathbf s} + H_0^{-1} v \hat{\mathbf s}) \nonumber \\
&\simeq& \frac{1}{2\sigma} \int\limits_{-\sigma}^{\sigma} dv 
\rho^{(r)}({\mathbf s} + H_0^{-1} v \hat{\mathbf s}) ~~~ .\\
\label{eqn:rhoz}
\end{eqnarray}
In this approximation, the effect of redshift space distortions leads to a
redistribution of the mass contained in an infinitesimally thin cylinder
of length $2 \sigma H_0^{-1}$ along the line of sight\footnote{We make this
  approximation in order to understand the qualitative aspects of density
  profile in redshift space and the corresponding features in the CPDF; the
  first of the two equations given here is used for the detailed calculations
  that follow.  Further, we use the NFW profile for our calculations
  and do not restrict ourselves to power law profiles.}.
Such a redistribution clearly leads to a smaller central density and
therefore the maximum density in a small ($R \ll r_{vir}$) finite
sized cell in redshift space is much smaller than its counterpart in
real space. 
However, the volume occupied by regions with this maximum density is much
larger and hence a large number of cells sample densities close to the
theoretical maximum. 
Therefore in redshift space, the CPDF should be steeper than its real
space counterpart.   
This is mainly because the volume occupied by densities close to the
theoretical maximum is much larger in redshift space than the corresponding
volume in real space. 

As an aside, we note that the density profile in the transverse direction
($s_\perp$) is shallower than the real space density profile.  
The difference in slope near the centre is unity for power law models. 
This dictates the shape of the power spectrum at wave numbers that are much
larger than $1/(\sigma H_0^{-1})$ and $1/r_{vir}$.  

Figure~3 shows the (unnormalised) probability distribution function for an
NFW halo computed using the approach followed by \citet{2003MNRAS.339..495T}
that we have outlined above.  
The probability is proportional to the volume occupied by regions with
the given density.   
This figure brings out the key feature of a smaller cutoff in the probability
distribution in redshift space.
As expected from the approximate treatment given above, the slope of the
non-Gaussian tail is also different: probability falls off very rapidly near
the maximum density. 
The real to redshift space mapping also enhances the probability at densities
below the maximum density in real space. 
This analysis is for a single halo and the net effect on the CPDF is
calculated by convolving the mass function with the density profile.
\begin{equation}
P(>N) \sim P(>\rho) \propto \int dM ~ n(M) ~ V(>\rho; M)
\end{equation}
here, $V(>\rho; M)$ is the volume occupied by regions with density greater than
$\rho$ in haloes of mass $M$.
The number density of haloes with mass between $M$ and $M + dM$ is given by
$n(M)~dM$. 
An analysis with finite sized cells requires a smoothing of the density
profile; however we find that we do not require significant smoothing
in order to explain the key features seen in N-Body simulations. 
The CPDF ($P(N)$) is obtained by differentiating $P(>N)$.  

The leading contribution to the tail of the CPDF comes from the most massive
haloes, as these haloes contribute the most to the volume above a
given density.  
The contribution from low mass haloes brings in small variation in the CPDF
due to the effect of smoothing, as compared to the CPDF for a single halo.  
This also implies that in our approach the relative forms of the non-Gaussian
tail of the CPDF in real space and redshift space do not depend strongly on
the power spectrum.    

The lower panel of Figure~3 shows the probability distribution function for
the $n=-2$ power law model computed using the halo model and assuming a NFW
profile in the limit of cell size much smaller than the virial radius of the
most massive haloes.  
In this limit the CPDF computed from the halo model closely resembles the CPDF
derived from a single halo shown in the top panel. 
In the following discussion, we shall compare these calculations with the CPDF
in N-Body simulations. 

There is a remarkable similarity in the distribution functions shown in
Figure~2 and Figure~3.
While the former is obtained from N-Body simulations, the latter is obtained
from the halo model.
The following assumptions have been used to compute the CPDF
using the halo model. 
\begin{itemize}
\item
The density profile of haloes is universal \citep{1996ApJ...462..563N}.
\item
The $1$D velocity dispersion $\sigma$ does not vary with distance from the
centre in a given halo.
\item
The cell size is much smaller than the virial radius of the most massive
haloes. 
\end{itemize} 

We find that the halo model predicts a form for the tail of the CPDF that is
remarkably similar to that seen in N-Body simulations. 

It is noteworthy that the CPDF for the $n=0$ model in redshift space stays
above the CPDF in real space over a much larger range of densities as compared
to the CPDFs for $n=-2$.  
In the analysis given above, we pointed out that the tail of the CPDF
computed using the halo model is dominated by the contribution from the most
massive haloes. 
Therefore the relative behaviour of the CPDF in real and redshift space at
high densities should have little dependence on the power spectrum as
long as the three assumptions listed above are valid.
We have used a cell size of $0.6$ grid lengths for computing the CPDF from
N-Body simulations, which is smaller than the typical radius of most massive
haloes by a factor of $4-5$ but not much more.   
Further, the most massive haloes for the $n=-2$ model are much smaller
than that for the $n=0$ model as we are able probe higher
non-linearities for the latter.  
This, in our view, is the reason for the differences seen in the two
panels of Figure~2. 

\begin{figure*}
\begin{tabular}{cccc}
\includegraphics[width=1.6in]{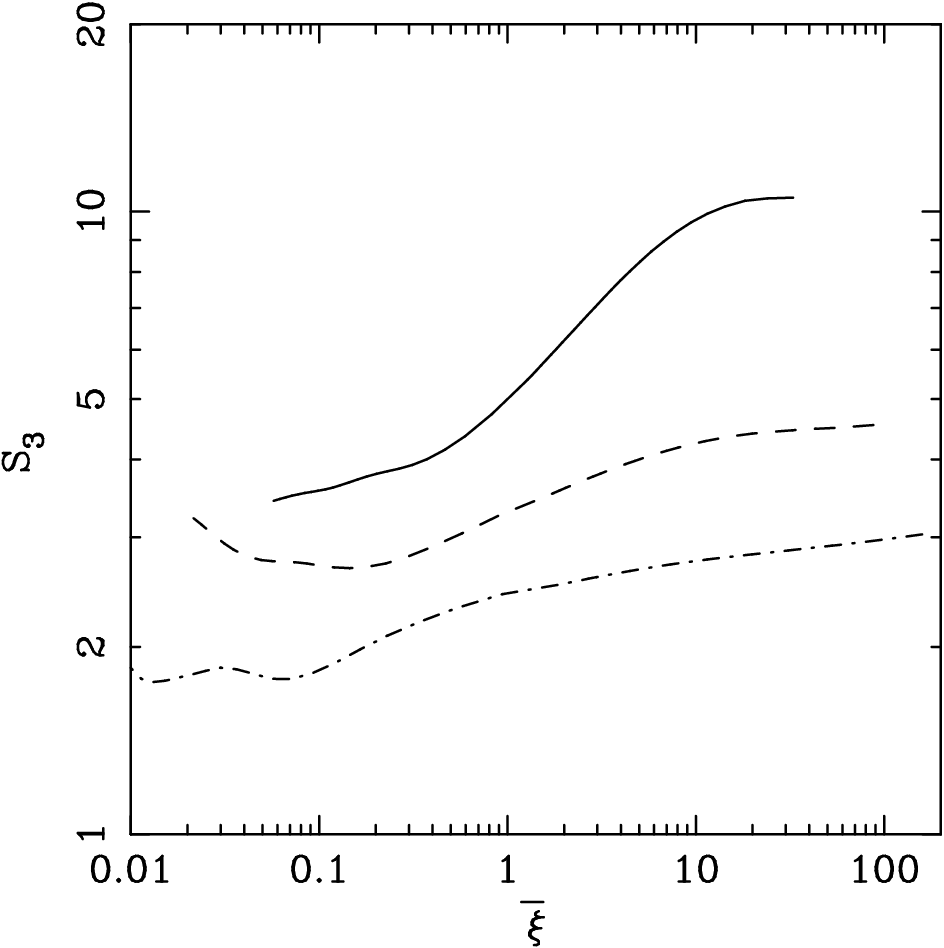}
&
\includegraphics[width=1.6in]{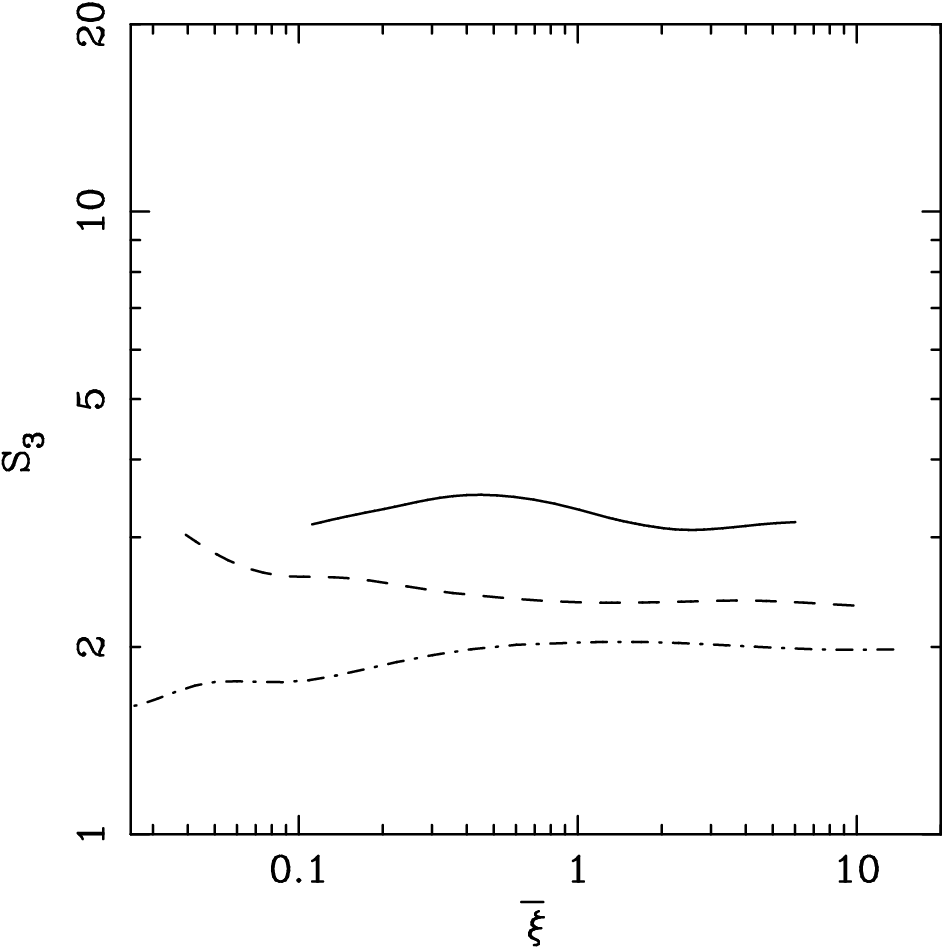}
&
\includegraphics[width=1.6in]{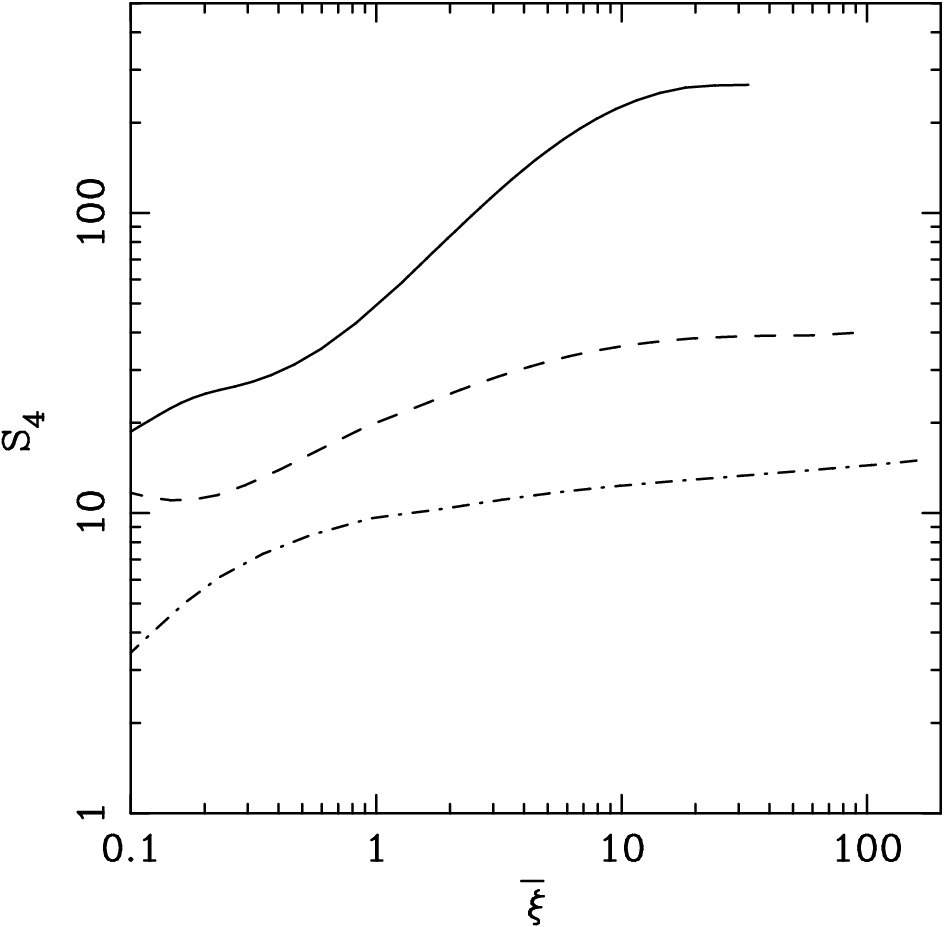}
&
\includegraphics[width=1.6in]{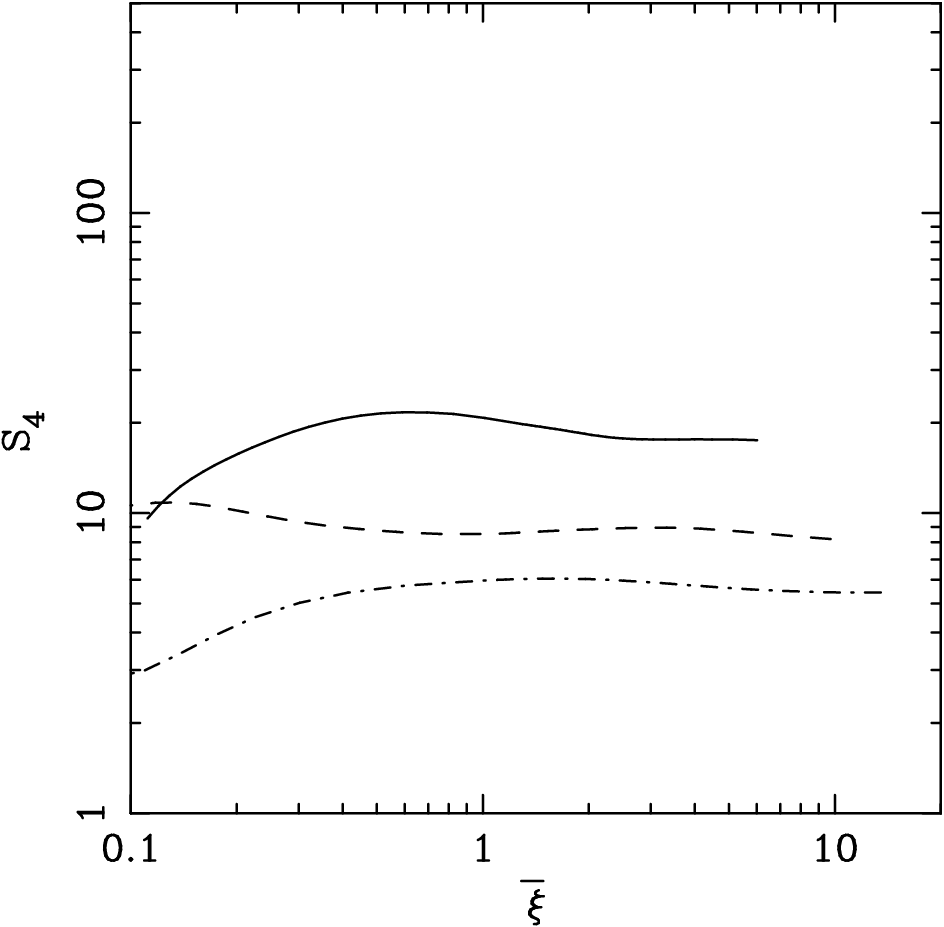}
\\
\includegraphics[width=1.6in]{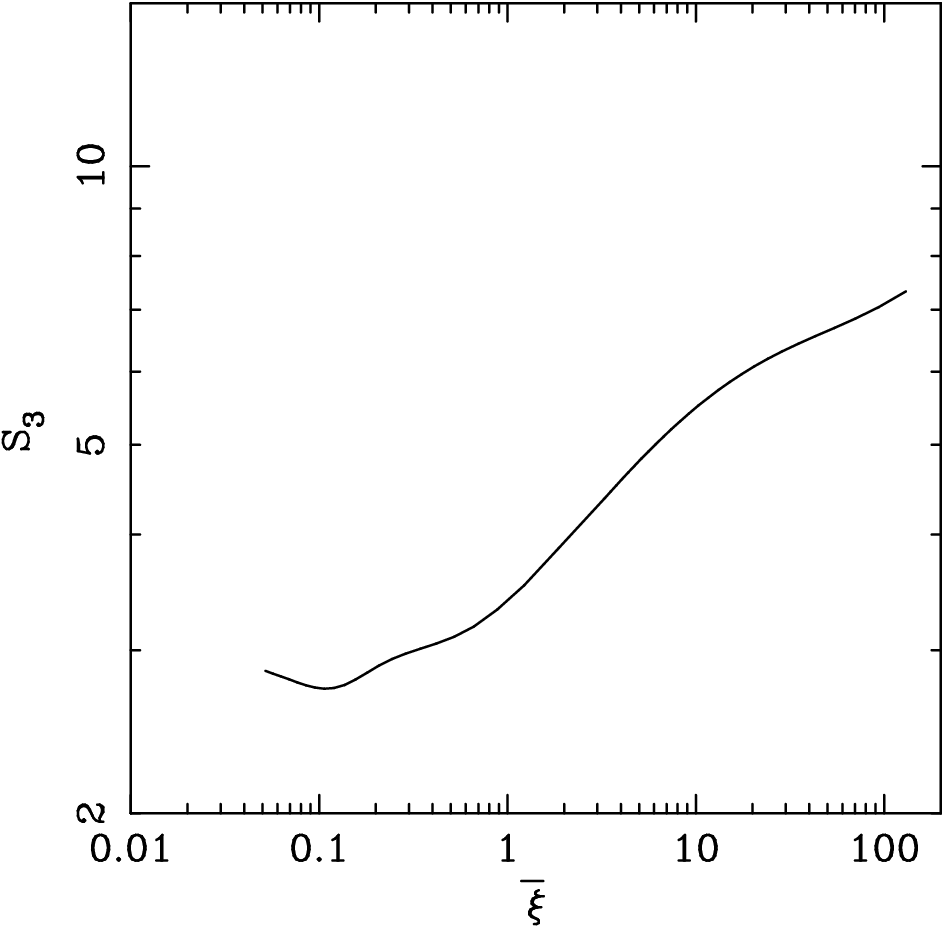}
&
\includegraphics[width=1.6in]{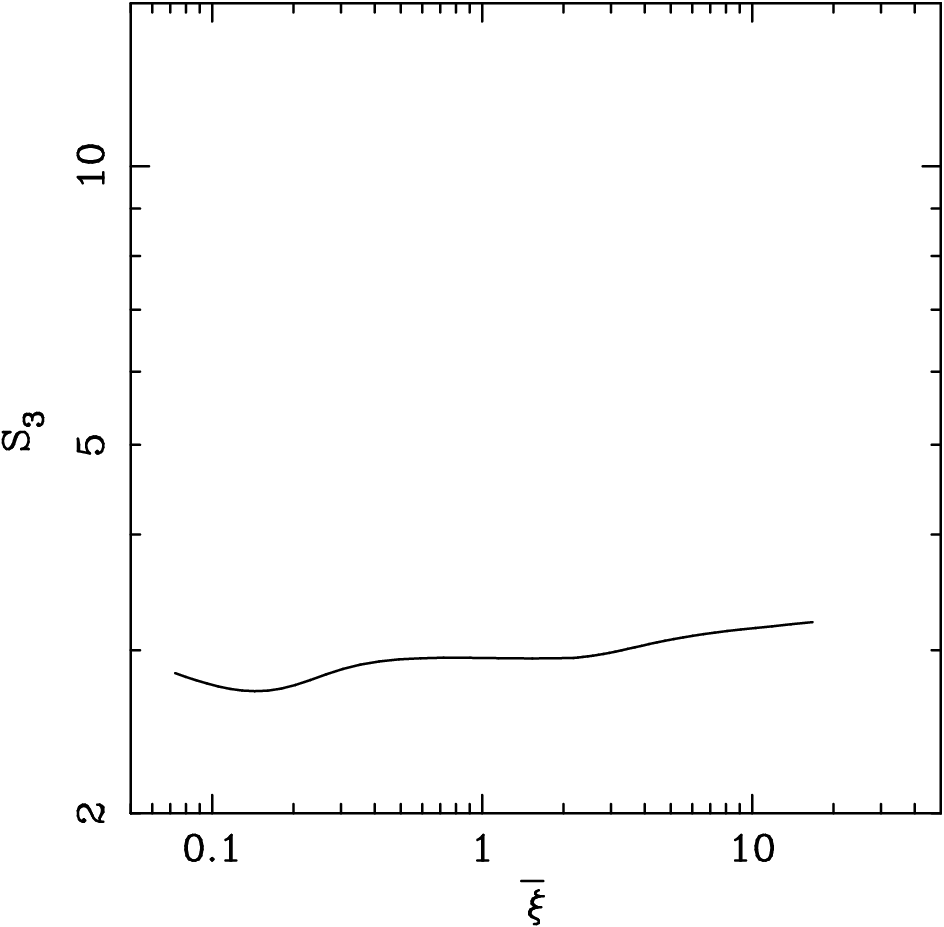}
&
\includegraphics[width=1.6in]{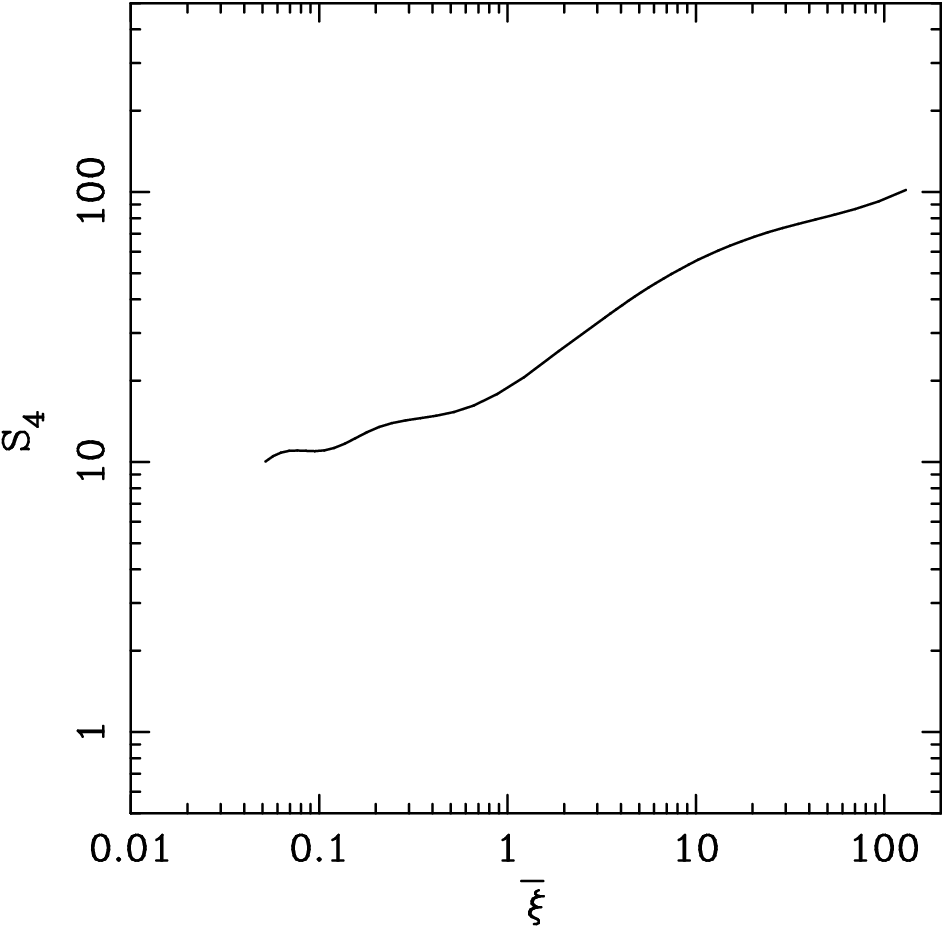}
&
\includegraphics[width=1.6in]{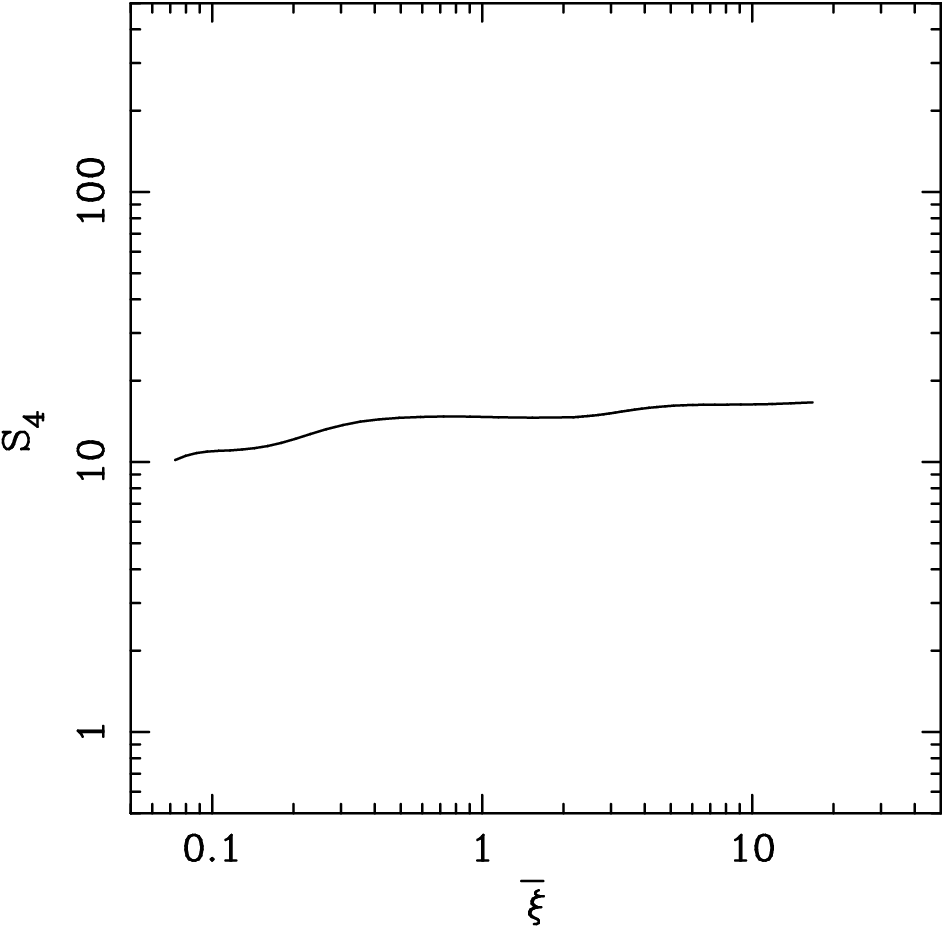}
\end{tabular}
\caption{This figure shows $S_3$ and $S_4$ as a function of $\bar\xi$.  The
  first column is for $S_3$ in real space, and the second column is for
  redshift space.  The third and fourth columns are for $S_4$ in real and
  redshift space respectively.  The top row is for power law models ($n=0$,
  $-1$ and $-2$) and the lower panel is for $\Lambda$CDM at $z=0$. In the top row, 
  solid, dashed and dot-dashed lines correspond to power law models with indices 
  $n=-2$, $n=-1$ and $n=0$ respectively. Same scale is used for plotting the moments 
  in real and redshift space to attract attention to the drastic reduction in spacing 
  between the models in the non-linear regime.}
\end{figure*}

\subsection{Moments of Counts in Cells}

The different shape of the high density tail of the CPDF in redshift space
has implications for the moments of counts in cells. 
The rapid fall in the CPDF for the distribution in redshift space implies a
much shorter non-Gaussian tail as compared to that in the real space. 
As the the tail contributes significantly to higher moments, we expect the
reduced moments to be much smaller in redshift space as compared to the real
space.   
Figure~4 shows $S_3$ and $S_4$ as a function of $\bar\xi$. 
These have been plotted for three power law models ($n=0$, $-1$ and $-2$) in
the top panel and for the $\Lambda$CDM model in the lower panel.  
We have plotted these quantities in real space as well as redshift space. 
For ease of comparison, the panels showing moments for the redshift space and
real space have been plotted with the same scale. 
Clearly, both the Skewness and Kurtosis are much smaller in redshift space as
compared to the corresponding values in real space.  
For the power law model with $n=-2$ and the $\Lambda$CDM model, the reduction
in values is significant.  
As the tail of CPDF becomes more and more relevant at scales with a large
amplitude of clustering (large $\bar\xi$), the reduction is most significant
at these scales. 
The ordering of values for $S_n$ for various models is preserved in the
mapping from real to redshift space, i.e., models with the larger $S_n$ in
real space also have a larger $S_n$ in redshift space. 
This applies only to power law models as the $\Lambda$CDM model does
not follow the ordering as we go from real to redshift space. 

A noteworthy feature is that the values of $S_n$ in redshift space
show very little scale dependence at non-linear scales ($\bar\xi \geq
1$).  
This again points to the finger-of-god effect as this effect is most
significant at non-linear scales.  

The significant reduction in the values of $S_n$ in redshift space in the
non-linear regime also results in reducing differences between different
models. 
As galaxy distribution is observed in the redshift space, this implies that
differentiating between variants of the $\Lambda$CDM model on the basis of
higher moments is much more difficult than would have been suspected on the
basis of the expected values in real space. 

In the regime of very small cell size, the differences in CPDF for different
models arise mainly due to different mass functions.  
Therefore we expect the differences in the CPDF for different models in real
space and redshift space to be of the same order, i.e., CPDF in real as well
as redshift space should vary by a similar amount with the power spectrum. 
However the sharp fall off for the CPDF in redshift space reduces the range
of densities over which the difference in CPDF contributes to the difference in
moments.  
This, we believe, is the reason behind much smaller differences in Skewness,
etc. for different models in redshift space as compared to the differences
between the same models in real space.  

\begin{figure*}
\begin{tabular}{cc}
\includegraphics[width=2.5in]{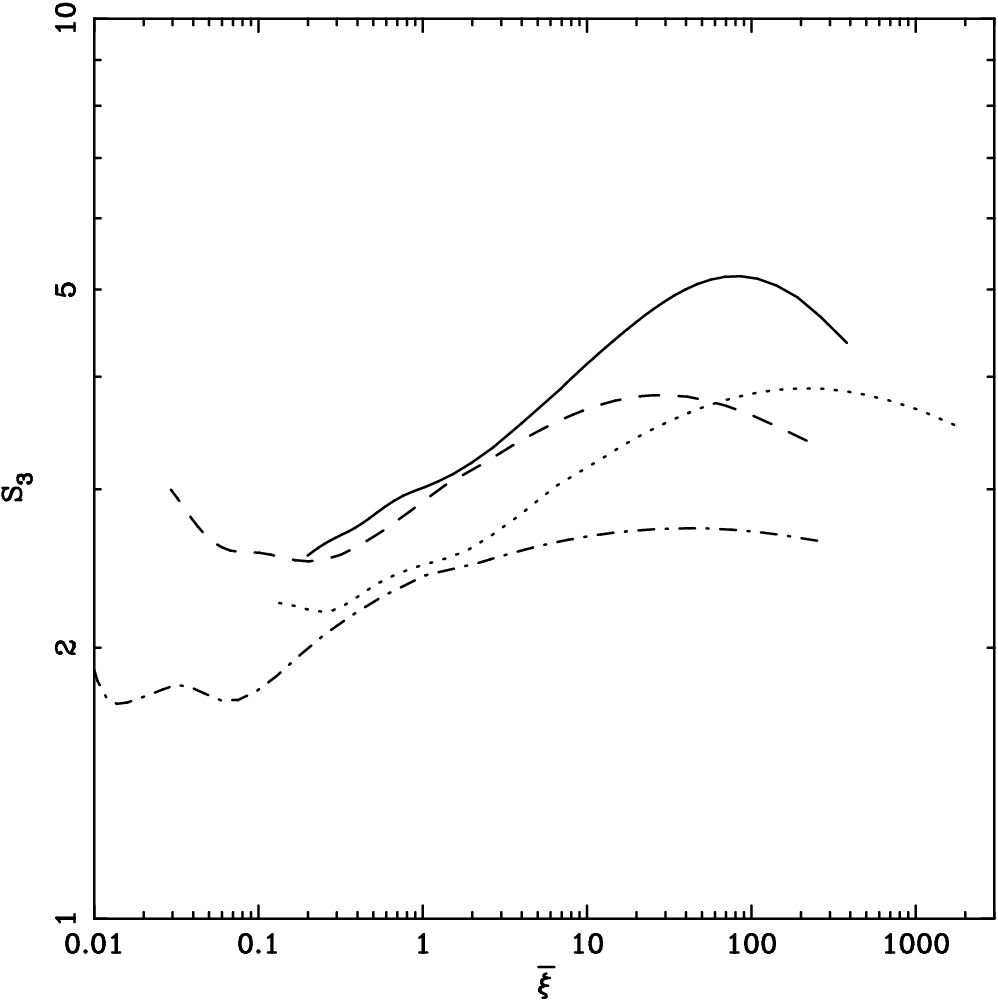} 
&
\includegraphics[width=2.5in]{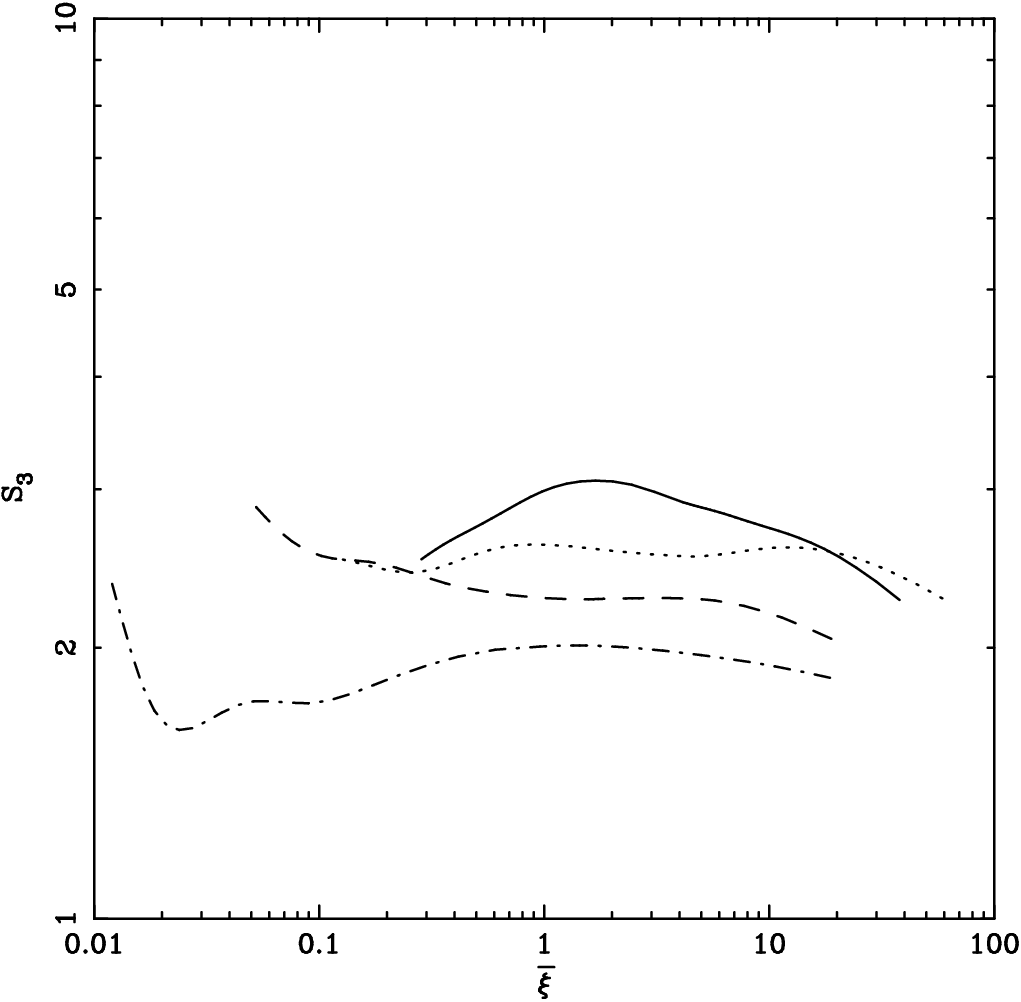}
\\
\includegraphics[width=2.5in]{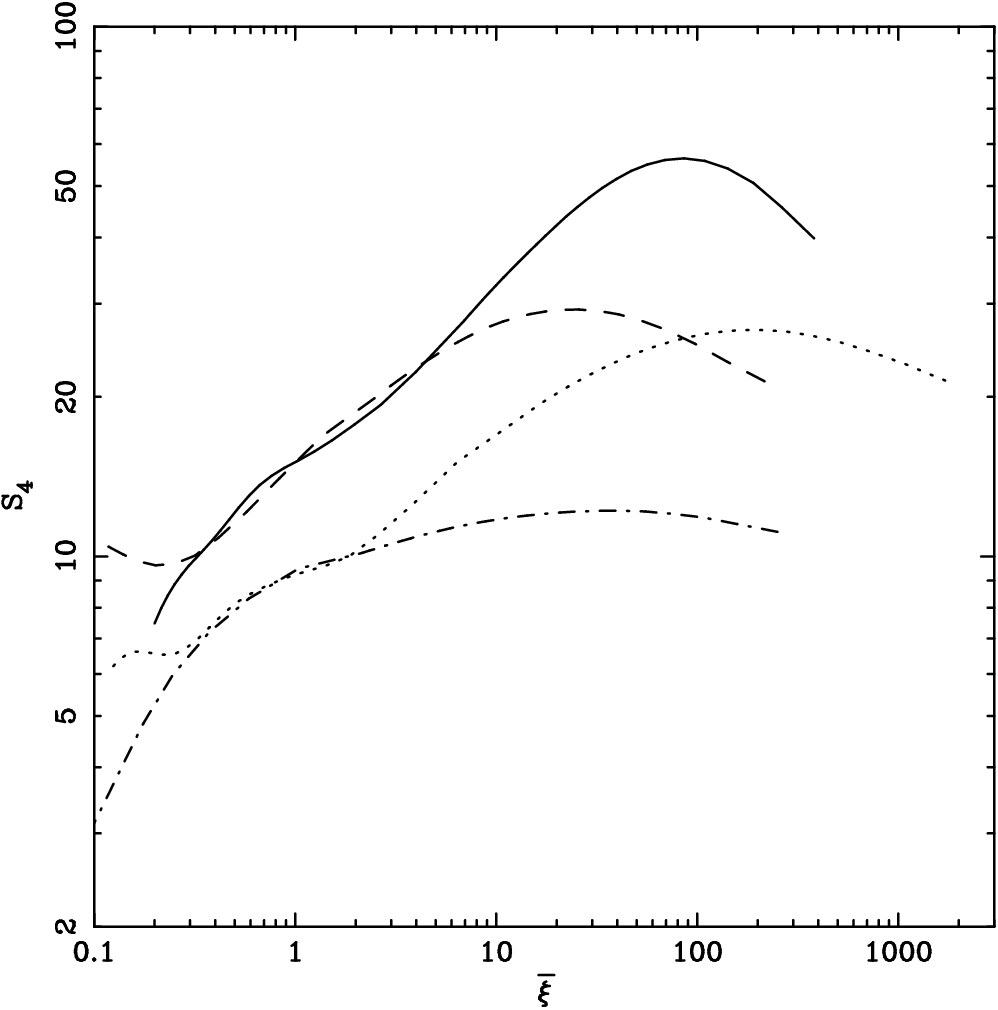}
&
\includegraphics[width=2.5in]{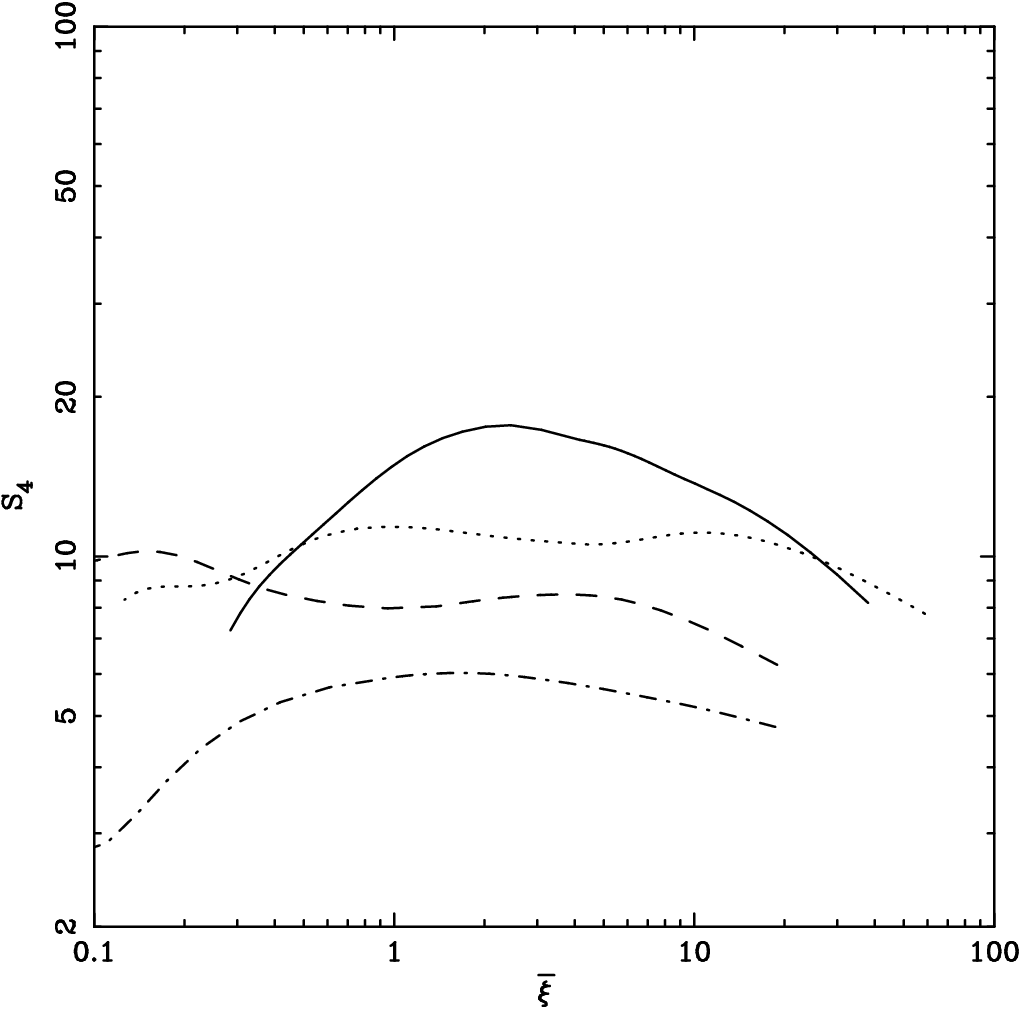}
\end{tabular}
\caption{$S_3$ (top row) and $S_4$ (lower row) have been plotted as a
  function of $\bar\xi$ for over dense regions (over density $\delta \ge 100$)
  in real space and redshift space. In all the panels solid, dashed
  and dot-dashed lines correspond to power law models with indices
  $n=-2$, $n=-1$ and $n=0$ respectively. The dotted line corresponds
  to the $\Lambda$CDM model.  The top left panel is for $S_3$
  in real space, the top right panel is for the same in redshift space.
  The lower left and right panels are for $S_4$ in real and redshift
  space, respectively.}  
\end{figure*}

Clustering properties of galaxies follow a pattern similar to clustering of
highly over dense regions.
Indeed, the generic evolution of galaxy correlation function can be deduced by
studying correlation function of over dense regions
\citep{1998MNRAS.299..417B}. 
Figure~5 shows $S_3$ and $S_4$ as a function of $\bar\xi$ for particles in
regions with $\delta \geq 100$. 
Here, the density contrast $\delta$ is computed in real space and only
particles in over dense regions with $\delta \geq 100$ are retained.  
We study the clustering properties of these particles in real and
redshift space and compute the two point correlation function and
reduced moments.  
We have plotted $S_3$ and $S_4$ for the three power law models and the
$\Lambda$CDM model in Figure~5.  
We see from the panels for real space that the moments are smaller for
over dense regions as compared to the distribution of all the
particles.  
Thus there are two reasons for reduction of moments in redshift space:
bias and redshift space distortions. 
Figure~5 suggests that redshift space distortions are often the more
dominant factor leading to reduction of Skewness, etc. 
At scales with $\bar\xi \geq 1$ the reduced moments are almost
constant; this feature is shared with the behaviour for the mass
distribution. 
The non-Gaussian tail of the CPDF gets contribution mainly from very massive
haloes, hence this is almost the same for the full distribution and over dense
regions. 
Therefore it is not surprising that the variation of Skewness,
etc. with scales is almost the same for matter distribution and over
dense regions. 
The variation of $S_n$ with $\bar\xi$ seen here is consistent with
that seen for galaxies in the two degree field galaxy redshift survey
\citep{2004MNRAS.352.1232C}.  
We are carrying out further analysis with models for occupation number
in haloes in order to do a more quantitative comparison of models with
the observed moments \citep{2001MNRAS.325.1039B,2006MNRAS.368...85T}.


\section{Discussion}

In this paper we have presented results of a study of the moments of counts of
cells in real space as well as redshift space for power law models.  
These models are used to develop insights into the variation of the
probability distribution function.  
We also use simulations of the $\Lambda$CDM model. 
We verify that in the weakly non-linear regime the Skewness and Kurtosis take
on values calculated using perturbation theory. 
This is true in both the real and redshift space, though the finger-of-god
effect leads to slightly smaller values in redshift space. 
We find that the values of higher moments like the Skewness and the
Kurtosis in the non-linear regime are very different in redshift space as
compared to the corresponding values in real space. 
Higher moments are very sensitive to the non-Gaussian tail of the CPDF and we
study that in order to understand the differences in values of moments
in real and redshift space. 
We find that at highly non-linear scales, the tail of the CPDF in redshift
space falls off very sharply much before the tail of CPDF in real space for
the same model. 
This pattern is similar for different power spectra and we hypothesise that
redshift space distortions (the finger-of-god effect) play a very important
role in determining the form of the CPDF at small scales.  
We use the halo model to compute the form of the non-Gaussian tail of the CPDF
in real and redshift space and find that we are able to reproduce the
behaviour observed in N-Body simulations. 

The rapid fall off of the CPDF in redshift space provides a natural
explanation for the much smaller values of the Skewness and Kurtosis in
the non-linear regime.  
Indeed, this also accounts for the much smaller differences in the values of
reduced moments for different models. 

We show that the above results carry over to the $\Lambda$CDM model as well. 
This indicates that at non-linear scales, higher moments of the CPDF are
relatively insensitive to the cosmological background. 

We also study moments for highly over dense regions in order to estimate the
effect of bias on Skewness, etc.   
In this case the moments in the non-linear regime take on smaller values as
compared to the values for all the particles at the same scales.  
The reduction of moments in real space is related to the value of bias and can 
be modelled analytically in the weakly non-linear regime
\citep{1997MNRAS.284..189M,1999ApJ...520...24D}.   
Thus there are two reasons for reduction of moments in the non-linear regime;
whereas bias applies in both the real space and redshift space, redshift
space distortions operate only in the latter. 
Figures~4 and 5 suggest that redshift space distortions are the stronger
effect in almost all the cases. 

We have shown that the halo model can be used successfully to model clustering
in redshift space. 
This is very different from other approaches that have been employed for
modelling the effect of redshift space distortions on clustering
\citep{2004PhRvD..70h3007S}. 
We propose to use this approach to model the three point correlation function,
etc. in future studies.  


\section*{Acknowledgements}

Numerical experiments for this study were carried out at the cluster computing
facility in the Harish-Chandra Research Institute
(http://cluster.mri.ernet.in).  
This research has made use of NASA's Astrophysics Data System.


\label{lastpage}


\begin{thebibliography}{}

\bibitem[\protect\citeauthoryear{Bagla}{1998a}]{1998MNRAS.297..251B} Bagla
  J.~S., 1998a, MNRAS, 297, 251  
 
\bibitem[\protect\citeauthoryear{Bagla}{1998b}]{1998MNRAS.299..417B} Bagla
  J.~S., 1998b, MNRAS, 299, 417 

\bibitem[\protect\citeauthoryear{{Bagla}}{2002}]{2002JApA...23..185B}{Bagla}
  J.~S.,  2002, Journal of Astrophysics and Astronomy, 23, 185
  (arXiv:astro-ph/9911025) 

\bibitem[\protect\citeauthoryear{Bagla}{2005}]{2004astro.ph.11043B} Bagla
  J.~S., 2005, Current Science 88, 1088 (arXiv:astro-ph/0411043) 

\bibitem[\protect\citeauthoryear{Bagla \&
    Padmanabhan}{1994}]{1994MNRAS.266..227B} Bagla J.~S., Padmanabhan T.,
  1994, MNRAS, 266, 227  

\bibitem[\protect\citeauthoryear{{Bagla} \&
    {Padmanabhan}}{1997a}]{1997MNRAS.286.1023B} {Bagla} J.~S.,  {Padmanabhan}
  T.,  1997a, MNRAS, 286, 1023  

\bibitem[\protect\citeauthoryear{Bagla and
  Padmanabhan}{1997b}]{1997Prama..49..161B} Bagla J. S. and Padmanabhan
  T. 1997b, Pramana -- Journal of Physics 49, 161 

\bibitem[\protect\citeauthoryear{{Bagla} \& {Ray}}{2003}]{2003NewA....8..665B}
  {Bagla} J.~S.,  {Ray} S.,  2003, New Astronomy, 8, 665 

\bibitem[\protect\citeauthoryear{Bagla \& Ray}{2005}]{2005MNRAS.358.1076B}
  Bagla J.~S., Ray S., 2005, MNRAS, 358, 1076

\bibitem[\protect\citeauthoryear{Bagla \& Prasad}{2006}]{2006astro.ph.01320B}
  Bagla J.~S., Prasad J., 2006, arXiv:astro-ph/0601320

\bibitem[\protect\citeauthoryear{Bardeen et al.}{1986}]{1986ApJ...304...15B}
  Bardeen J.~M., Bond J.~R., Kaiser N., Szalay A.~S., 1986, ApJ, 304, 15  
 
\bibitem[\protect\citeauthoryear{Benson}{2001}]{2001MNRAS.325.1039B} Benson 
A.~J., 2001, MNRAS, 325, 1039

\bibitem[\protect\citeauthoryear{{Bernardeau}
    et~al.}{2002}]{2002PhR...367....1B} {Bernardeau} F.,  {Colombi} S.,
  {Gazta{\~ n}aga} E.,    {Scoccimarro} R., 2002, Physics Reports, 367, 1  

\bibitem[\protect\citeauthoryear{Bertschinger \&
    Dekel}{1989}]{1989ApJ...336L...5B} Bertschinger E., Dekel A., 1989, ApJ,
  336, L5  
 
\bibitem[\protect\citeauthoryear{Bertschinger}{1998}]{1998ARA&A..36..599B}
  Bertschinger E., 1998, ARA\&A, 36, 599  

\bibitem[\protect\citeauthoryear{Binney}{1977}]{1977ApJ...215..483B} Binney
  J., 1977, ApJ, 215, 483  

\bibitem[\protect\citeauthoryear{Brenier et al.}{2003}]{2003MNRAS.346..501B}
  Brenier Y., Frisch U., H{\' e}non M., Loeper G., Matarrese S., Mohayaee R.,
  Sobolevski{\u i} A., 2003, MNRAS, 346, 501  
 
\bibitem[\protect\citeauthoryear{Brainerd, Scherrer, \&
    Villumsen}{1993}]{1993ApJ...418..570B} Brainerd T.~G., Scherrer R.~J.,
  Villumsen J.~V., 1993, ApJ, 418, 570  

\bibitem[\protect\citeauthoryear{Brainerd \&
    Villumsen}{1994}]{1994ApJ...431..477B} Brainerd T.~G., Villumsen J.~V.,
  1994, ApJ, 431, 477  

\bibitem[\protect\citeauthoryear{Branchini, Eldar, \&
    Nusser}{2002}]{2002MNRAS.335...53B} Branchini E., Eldar A., Nusser A.,
  2002, MNRAS, 335, 53  
 
\bibitem[\protect\citeauthoryear{Buchert \&
    Dominguez}{1998}]{1998A&A...335..395B} Buchert T., Dominguez A., 1998,
  A\&A, 335, 395  
 
\bibitem[\protect\citeauthoryear{{Colombi}
  et~al.}{1994}]{1994A&A...281..301C} {Colombi} S.,  {Bouchet} F.~R.,
  {Schaeffer} R.,  1994, A\&A, 281, 301 

\bibitem[\protect\citeauthoryear{Colombi, Bouchet, \&
    Schaeffer}{1995}]{1995ApJS...96..401C} Colombi S., Bouchet F.~R.,
  Schaeffer R., 1995, ApJS, 96, 401  

\bibitem[\protect\citeauthoryear{Colombi, Bouchet, \&
    Hernquist}{1996}]{1996ApJ...465...14C} Colombi S., Bouchet F.~R.,
  Hernquist L., 1996, ApJ, 465, 14  
 
\bibitem[\protect\citeauthoryear{Cooray \& Sheth}{2002}]{2002PhR...372....1C}
  Cooray A., Sheth R., 2002, PhR, 372, 1  

\bibitem[\protect\citeauthoryear{Croft \&
    Gaztanaga}{1997}]{1997MNRAS.285..793C} Croft R.~A.~C., Gaztanaga E., 1997,
    MNRAS, 285, 793  
 
\bibitem[\protect\citeauthoryear{Croton et al.}{2004}]{2004MNRAS.352.1232C}
  Croton D.~J., et al., 2004, MNRAS, 352, 1232  

\bibitem[\protect\citeauthoryear{Davis \& Peebles}{1977}]{1977ApJS...34..425D}
  Davis M., Peebles P.~J.~E., 1977, ApJS, 34, 425  

\bibitem[\protect\citeauthoryear{Dekel, Bertschinger, \&
    Faber}{1990}]{1990ApJ...364..349D} Dekel A., Bertschinger E., Faber S.~M.,
  1990, ApJ, 364, 349  
 
\bibitem[\protect\citeauthoryear{Dekel \& Lahav}{1999}]{1999ApJ...520...24D}
  Dekel A., Lahav O., 1999, ApJ, 520, 24  

\bibitem[\protect\citeauthoryear{Frisch et al.}{2002}]{2002Natur.417..260F}
  Frisch U., Matarrese S., Mohayaee R., Sobolevski A., 2002, Natur, 417, 260  
 
\bibitem[\protect\citeauthoryear{Fry}{1996}]{1996ApJ...461L..65F} Fry J.~N.,
  1996, ApJ, 461, L65  

\bibitem[\protect\citeauthoryear{Gurbatov, Saichev, \&
    Shandarin}{1989}]{1989MNRAS.236..385G} Gurbatov S.~N., Saichev A.~I.,
  Shandarin S.~F., 1989, MNRAS, 236, 385  

\bibitem[\protect\citeauthoryear{Hamilton et al.}{1991}]{1991ApJ...374L...1H}
  Hamilton A.~J.~S., Kumar P., Lu E., Matthews A., 1991, ApJ, 374, L1  
 
\bibitem[\protect\citeauthoryear{Hoyle}{1953}]{1953ApJ...118..513H} Hoyle F.,
  1953, ApJ, 118, 513  

\bibitem[\protect\citeauthoryear{Hui \&
    Bertschinger}{1996}]{1996ApJ...471....1H} Hui L., Bertschinger E., 1996,
  ApJ, 471, 1  
 
\bibitem[\protect\citeauthoryear{Hwang \& Noh}{2005}]{2005astro.ph..7159H}
  Hwang J., Noh H., 2005, astro, arXiv:astro-ph/0507159  

\bibitem[\protect\citeauthoryear{Jain, Mo, \&
    White}{1995}]{1995MNRAS.276L..25J} Jain B., Mo H.~J., White S.~D.~M.,
  1995, MNRAS, 276, L25  
 
\bibitem[\protect\citeauthoryear{Jones et al.}{2005}]{2005RvMP...76.1211J}
  Jones B.~J., Mart{\'{\i}}nez V.~J., Saar E., Trimble V., 2005, RvMP, 76,
  1211  

\bibitem[\protect\citeauthoryear{Kaiser}{1984}]{1984ApJ...284L...9K} Kaiser
  N., 1984, ApJ, 284, L9  

\bibitem[\protect\citeauthoryear{Kaiser}{1987}]{1987MNRAS.227....1K} Kaiser
  N., 1987, MNRAS, 227, 1  
 
\bibitem[\protect\citeauthoryear{{Kanekar}}{2000}]{2000ApJ...531...17Ka}
  {Kanekar} N.,  2000, ApJ, 531, 17 

\bibitem[\protect\citeauthoryear{{Ma}}{1998}]{1998ApJ...508L...5M} {Ma} C.,
  1998, ApJL, 508, L5 

\bibitem[\protect\citeauthoryear{Matarrese et al.}{1992}]{1992MNRAS.259..437M}
  Matarrese S., Lucchin F., Moscardini L., Saez D., 1992, MNRAS, 259, 437  
 
\bibitem[\protect\citeauthoryear{Matsubara}{1994}]{1994ApJ...424...30M}
  Matsubara T., 1994, ApJ, 424, 30 

\bibitem[\protect\citeauthoryear{Matarrese et al.}{1997}]{1997MNRAS.286..115M}
  Matarrese S., Coles P., Lucchin F., Moscardini L., 1997, MNRAS, 286, 115  

\bibitem[\protect\citeauthoryear{Mo \& White}{1996}]{1996MNRAS.282..347M} Mo
  H.~J., White S.~D.~M., 1996, MNRAS, 282, 347  

\bibitem[\protect\citeauthoryear{Mo, Jing, \&
    White}{1997}]{1997MNRAS.284..189M} Mo H.~J., Jing Y.~P., White S.~D.~M.,
  1997, MNRAS, 284, 189 

\bibitem[\protect\citeauthoryear{Mohayaee et al.}{2003}]{2003A&A...406..393M}
  Mohayaee R., Frisch U., Matarrese S., Sobolevskii A., 2003, A\&A, 406, 393  
 
\bibitem[\protect\citeauthoryear{Mohayaee, Tully, \&
    Frisch}{2004}]{2004astro.ph.10063M} Mohayaee R., Tully B., Frisch U.,
  2004, arXiv:astro-ph/0410063  
 
\bibitem[\protect\citeauthoryear{Monaco \&
    Efstathiou}{1999}]{1999MNRAS.308..763M} Monaco P., Efstathiou G., 1999,
  MNRAS, 308, 763  
 
\bibitem[\protect\citeauthoryear{Narayanan \&
    Weinberg}{1998}]{1998ApJ...508..440N} Narayanan V.~K., Weinberg D.~H.,
  1998, ApJ, 508, 440  
 
\bibitem[\protect\citeauthoryear{Narayanan \&
    Croft}{1999}]{1999ApJ...515..471N} Narayanan V.~K., Croft R.~A.~C., 1999,
  ApJ, 515, 471  
 
\bibitem[\protect\citeauthoryear{Navarro, Frenk, \& White}
  {1996}]{1996ApJ...462..563N} Navarro J.~F., Frenk C.~S., White S.~D.~M.,
  1996, ApJ 462, 563

\bibitem[\protect\citeauthoryear{Nityananda \&
    Padmanabhan}{1994}]{1994MNRAS.271..976N} Nityananda R., Padmanabhan T.,
  1994, MNRAS, 271, 976 
 
\bibitem[\protect\citeauthoryear{Nusser \& Dekel}{1992}]{1992ApJ...391..443N}
  Nusser A., Dekel A., 1992, ApJ, 391, 443  
 
\bibitem[\protect\citeauthoryear{Padmanabhan}{1996}]{1996MNRAS.278L..29P}
  Padmanabhan T., 1996, MNRAS, 278, L29  
 
\bibitem[\protect\citeauthoryear{{Padmanabhan}
    et~al.}{1996}]{1996ApJ...466..604P} {Padmanabhan} T.,  {Cen} R.,
    {Ostriker} J.~P.,    {Summers} F.~J.,  1996, ApJ, 466, 604 

\bibitem[\protect\citeauthoryear{{Padmanabhan}}{2002}]{2002tagc.book.....P}
  {Padmanabhan} T.,  2002, {Theoretical Astrophysics, Volume III: Galaxies and
  Cosmology}. Cambridge University Press. 

\bibitem[\protect\citeauthoryear{Peacock \& Dodds}{1996}]{1996MNRAS.280L..19P}
  Peacock J.~A., Dodds S.~J., 1996, MNRAS, 280, L19  
 
\bibitem[\protect\citeauthoryear{{Peacock}}{1999}]{1999coph.book.....P}
  {Peacock} J.~A.,  1999, Cosmological physics. Cambridge University Press. 

\bibitem[\protect\citeauthoryear{Peebles}{1974}]{1974A&A....32..391P} Peebles
  P.~J.~E., 1974, A\&A, 32, 391  
 
\bibitem[\protect\citeauthoryear{{Peebles}}{1980}]{1980lssu.book.....P}
  {Peebles} P.~J.~E.,  1980, {The large-scale structure of the universe}.
  Princeton University Press, Princeton. 

\bibitem[\protect\citeauthoryear{Peebles}{1985}]{1985ApJ...297..350P} Peebles
  P.~J.~E., 1985, ApJ, 297, 350  
 
\bibitem[\protect\citeauthoryear{{Ray} \&
    {Bagla}}{2004}]{2004astro.ph..5220Ra} {Ray} S.,  {Bagla} J.~S.,  2004,
    astro-ph/0405220 

\bibitem[\protect\citeauthoryear{Rees \& Ostriker}{1977}]{1977MNRAS.179..541R}
  Rees M.~J., Ostriker J.~P., 1977, MNRAS, 179, 541  

\bibitem[\protect\citeauthoryear{Roukema et al.}{1999}]{1999MNRAS.305..151R}
  Roukema B.~F., Valls-Gabaud D., Mobasher B., Bajtlik S., 1999, MNRAS, 305,
  151  

\bibitem[\protect\citeauthoryear{Sahni \& Coles}{1995}]{1995PhR...262....1S}
  Sahni V., Coles P., 1995, PhR, 262, 1  
 
\bibitem[\protect\citeauthoryear{Scoccimarro}{2004}]{2004PhRvD..70h3007S} 
Scoccimarro R., 2004, PhRvD, 70, 083007 

\bibitem[\protect\citeauthoryear{Seljak}{2001}]{2001MNRAS.325.1359S} Seljak
  U., 2001, MNRAS, 325, 1359  

\bibitem[\protect\citeauthoryear{Shandarin \&
    Zeldovich}{1989}]{1989RvMP...61..185S} Shandarin S.~F., Zeldovich Y.~B.,
    1989, RvMP, 61, 185  
 
\bibitem[\protect\citeauthoryear{Sheth \& Tormen}{2004}]{2004MNRAS.350.1385S}
  Sheth R.~K., Tormen G., 2004, MNRAS, 350, 1385 

\bibitem[\protect\citeauthoryear{Sheth, Mo, \&
    Tormen}{2001}]{2001MNRAS.323....1S} Sheth R.~K., Mo H.~J., Tormen G.,
  2001, MNRAS, 323, 1  

\bibitem[\protect\citeauthoryear{Silk}{1977}]{1977ApJ...211..638S} Silk J.,
  1977, ApJ, 211, 638  

\bibitem[\protect\citeauthoryear{Smith et al.}{2003}]{2003MNRAS.341.1311S}
  Smith R.~E., et al., 2003, MNRAS, 341, 1311  
 
\bibitem[\protect\citeauthoryear{Springel, Yoshida, \&
    White}{2001}]{2001NewA....6...79S} Springel V., Yoshida N., White
  S.~D.~M., 2001, NewA, 6, 79  

\bibitem[\protect\citeauthoryear{Taruya, Hamana, \&
    Kayo}{2003}]{2003MNRAS.339..495T} Taruya A., Hamana T., Kayo I., 2003,
  MNRAS, 339, 495  
 
\bibitem[\protect\citeauthoryear{Taylor \&
    Valentine}{1999}]{1999MNRAS.306..491T} Taylor A., Valentine H., 1999,
  MNRAS, 306, 491  
 
\bibitem[\protect\citeauthoryear{Taylor \& Watts}{2000}]{2000MNRAS.314...92T}
  Taylor A.~N., Watts P.~I.~R., 2000, MNRAS, 314, 92  
 
\bibitem[\protect\citeauthoryear{Tegmark \&
    Peebles}{1998}]{1998ApJ...500L..79T} Tegmark M., Peebles P.~J.~E., 1998,
  ApJ, 500, L79  

\bibitem[\protect\citeauthoryear{Tinker, Weinberg, \& 
Zheng}{2006}]{2006MNRAS.368...85T} Tinker J.~L., Weinberg D.~H., Zheng Z., 
2006, MNRAS, 368, 85

\bibitem[\protect\citeauthoryear{Watts \& Taylor}{2001}]{2001MNRAS.320..139W}
  Watts P.~I.~R., Taylor A.~N., 2001, MNRAS, 320, 139  
 
\bibitem[\protect\citeauthoryear{Weinberg}{1992}]{1992MNRAS.254..315W}
  Weinberg D.~H., 1992, MNRAS, 254, 315  
 
\bibitem[\protect\citeauthoryear{White}{2001}]{2001MNRAS.321....1W} White M.,
  2001, MNRAS, 321, 1  

\bibitem[\protect\citeauthoryear{Whiting}{2000}]{2000ApJ...533...50W} Whiting
  A.~B., 2000, ApJ, 533, 50  

\bibitem[\protect\citeauthoryear{Zel'Dovich}{1970}]{1970A&A.....5...84Z}
  Zel'Dovich Y.~B., 1970, A\&A, 5, 84  

\end{thebibliography}
\end{document}